\documentclass[12pt,preprint]{emulateapj}
\usepackage{graphicx}
\usepackage{epsfig,rotating,amsmath,natbib,lscape}
\nocite{*}

\slugcomment{ApJ; received 2008 September; published 2009 October}

\shorttitle{Disruption and Fueling of M33}
\shortauthors{Putman et al.}

\def\gtrapprox{\;\lower 0.5ex\hbox{$\buildrel >\over \sim\ $}}
\def\lessapprox{\;\lower 0.5ex\hbox{$\buildrel < \over \sim\ $}}

\def\Msun  {${\rm M}_\odot$}
\def\deg   {$^\circ$}

\def\kms   {km s$^{-1}$}

\def\cm   {cm$^{-2}$}

\begin{document}

\title{The Disruption and Fueling of M33}

\author{M.E. Putman\altaffilmark{1}, J.E.G. Peek\altaffilmark{2}, A. Muratov\altaffilmark{3}, O.Y. Gnedin\altaffilmark{3}, W. Hsu\altaffilmark{3}, K.A. Douglas\altaffilmark{4,5}, C. Heiles\altaffilmark{3}, S. Stanimirovic\altaffilmark{6}, E.J. Korpela\altaffilmark{4}, S.J. Gibson\altaffilmark{7,8}}
\altaffiltext{1}{Department of Astronomy, Columbia University, New York, NY 10027}
\altaffiltext{2}{Astronomy Department, University of California at Berkeley, Berkeley, CA 94720-3411}
\altaffiltext{3}{Department of Astronomy, University of Michigan, Ann Arbor, MI 48109-1042}
\altaffiltext{4}{Space Sciences Laboratory, University of California, Berkeley, CA 94720}
 \altaffiltext{5}{School of Physics, University of Exeter, Stocker Road, Exeter, United Kingdom EX4 4QL}
\altaffiltext{6}{Department of Astronomy, University of Wisconsin - Madison, Madison, WI 53706-1582}
\altaffiltext{7}{Arecibo Observatory, National Astronomy and Ionosphere Center, Arecibo, PR 00612}
 \altaffiltext{8}{Department of Physics and Astronomy, Western Kentucky University, Bowling Green, KY 42101}
\begin{abstract}

The disruption of the M33 galaxy is evident from its extended gaseous structure.
We present new data from the Galactic Arecibo L-Band Feed Array HI (GALFA-HI) Survey that show the full
extent and detailed spatial and kinematic structure of M33's neutral hydrogen.
Over 18\% of the HI mass of M33 (M$_{\rm HI_{tot}}=1.4 \times 10^9$ \Msun) is found beyond the star forming disk as traced in the far-ultraviolet (FUV).  The most distinct features are extended warps, an arc from the northern warp to the disk, diffuse gas surrounding the galaxy, and a southern cloud with a filament back to the galaxy.   The features extend out to 22 kpc from the galaxy center (18 kpc from the edge of the FUV disk) and the gas is directly connected to M33's disk.  Only five discrete clouds (i.e., gas not directly connected to M33 in position-velocity space) are catalogued in the vicinity of M33, and these clouds show similar properties to Galactic and M31 halo clouds.
M33's gaseous features most likely originate from the tidal disruption of M33 by M31 1-3 Gyr ago as shown from an orbit analysis which results in a tidal radius $<15$ kpc in the majority of M33's possible orbits.  M33 is now beyond the disruptive gravitational influence of M31 and the gas appears
  to be returning to M33's disk and redistributing its star formation fuel.  
  M33's high mean velocity dispersion in the disk ($\sim$18.5 \kms) may also be consistent with the previous interaction and high rate of star formation.
  M33 will either exhaust its star formation fuel in the next few Gyrs or eventually become star formation fuel for M31.  The latter represents the accretion of a large gaseous satellite by a spiral galaxy, similar to the Magellanic Clouds' relationship to the Galaxy.  
 \end{abstract}


\section{Introduction}

Large spiral galaxies at low redshift are thought to gradually evolve through the accretion of material from their halos.  This halo
material originates from the accretion of satellite galaxies and the cooling of the hot gaseous halo, with the hot halo most likely originating from a combination of the initial baryon collapse and galaxy feedback.   Continual 
gas accretion provides star formation fuel and
allows for the range of stellar populations and metallicities we find in galaxies.
Simulations that examine the methods galaxies obtain gas indicate the dominant mechanism depends 
on a galaxy's total mass and redshift \citep{keres09, keres05, brooks09}.  At $z=0$ massive galaxies are  expected to accrete gas through satellites and the gradual accretion of hot halo gas, while less massive galaxies ($<$ few $\times~10^{11}$ \Msun) obtain
star formation fuel primarily via direct accretion of cold gas along cosmic filaments.

 Observational evidence for the various accretion modes remains limited.  The Galaxy and M31 have been shown to
have numerous halo clouds (e.g., Muller et al. 1963; Putman et al. 2002; Thilker et al. 2004\nocite{muller63, putman02, thilker04}), with their origin definitely partly from 
satellite accretion \citep{mathewson74, putman03}, and possibly also from condensing halo clouds in the hot
halo medium \citep{peek08,maller04}.  There has been additional evidence for halo clouds that will
fuel spiral galaxies through deep HI observations of nearby spirals \citep{oosterloo07, fraternali01}.  In all of these cases, the
amount of fuel currently found in the halo is not large considering the expected star formation rates and
chemical evolution models.  
For instance, the Milky Way is expected to have an average infall of 1 \Msun/yr \citep{chiappini01, fenner03}, but this reservoir is not readily apparent from either
the accretion of gas from small satellites \citep{grcevich09} or the condensing halo clouds \citep{peek08}.  
In the case of galaxies smaller than the Milky Way, observational evidence for halo clouds remains limited, and it is possible they will largely not be found, as the bulk of the accretion may have 
been completed for these systems at early times.

M33 is often considered a dwarf spiral galaxy of the Local Group, with a total mass of $\sim5\times10^{10}$ \Msun (Corbelli 2003\nocite{corbelli03}), which is $\sim$20 times lower
than the Milky Way and M31, and only $\sim$2 times higher than the combined mass of the Magellanic Clouds.   
M33 is $\sim730$ kpc from the Milky Way and $\sim$ 200 kpc from M31 \citep{brunthaler05}.
Its stellar component indicates it is a fairly quiescent galaxy with no evidence for a recent interaction.
Deep optical surveys have found a stellar halo component, but no clear evidence for the stream-like
sub-structures found in the halo of the Milky Way and M31 (Ferguson et al. 2007; McConnachie et al. 2006\nocite{ferguson07, mcconnachie06}).
In neutral hydrogen there is evidence for a strong warp that has been examined by \cite{corbelli97} (hereafter CS97) and Rogstad, Wright \& Lockhart (1976).  Both authors concluded that
the tidal force from M31 is unlikely to be sufficient to generate the warp.   On the other hand, recent
work estimating the motion of M31 from indirect methods indicates that M33 and M31 are
almost certainly gravitationally bound and M33 should show signatures of tidal perterbation \citep{vandermarel08,loeb05}.

Sensitive large area neutral hydrogen surveys of M33's extended halo can investigate the methods that M33 is
losing and/or obtaining star formation fuel, and the fate of this galaxy in the context of galaxy mergers at $z=0$.
Recent observations taken as part of the GALFA-HI Survey provide
these data at unprecedented resolution (3.4\arcmin, 0.18 \kms) and sensitivity (0.1 K ($3\sigma$) over 5 \kms).\footnote{The data are publicly available through a link at http://sites.google.com/site/galfahi/.}  See also Grossi et al. (2008\nocite{grossi08}) for data taken with a different spectrometer and reduced with different techniques as part of the ALFALFA survey.
In this paper we present the GALFA-HI Survey observations of the HI environment of M33, complete simulations of the orbit of M33 relative to M31, and discuss the results
in the context of fueling both M33 and M31.  The gaseous environment of M33 is an important tracer of the past and future evolution of the galaxies of the Local Group.

\section{Observations and Data Reduction}

The HI data presented here were obtained with the Galactic Arecibo L-Band Feed Array (GALFA) spectrometer
on the Arecibo Radio Telescope.   The Arecibo L-Band Feed Array (ALFA) is a 7-beam array of receivers similar in design to the 13-beam array used for the HI Parkes All-Sky Survey (HIPASS; Barnes et al. 2001\nocite{barnes01}).
GALFA data provide a channel width of 0.18 \kms~and cover a velocity range of $-753$ to $+753$ \kms~ in the Local Standard of Rest (LSR) frame.   The spatial resolution is approximately 3.4\arcmin.  At the distance to M33 (730 kpc), 3.4\arcmin~corresponds to 720 parsecs.  The GALFA-HI Survey will eventually map the Arecibo sky from Decl. $= -1 - 38$\deg~and the full RA range.  The 
data presented here are from the TOGS (Turn On GALFA Survey) portion of the GALFA-HI Survey.  
TOGS obtains data by commensally observing with the
extragalactic HI surveys (ALFALFA (e.g., Giovanelli et al. 2005\nocite{giovanelli05}) and AGES (e.g., Auld et al. 2006\nocite{auld06})), with the majority of the data shown here
taken commensally with ALFALFA.  The observing
method used is drift scans with our least-squares frequency switching calibration procedure run at the start of each observing run.   There are also some basketweave scans (driving the telescope only in zenith) taken as part of the A2050 GALFA proposal (PI's:  Peek \& Heiles) integrated into the final data cube that improve any drift scan artifacts and increase the sensitivity.

The reduction process and data are discussed in Peek \& Heiles (2008\nocite{peekheiles}) 
and include corrections for the IF bandpass and gain variation in the receivers.  The dataset shown here includes a sidelobe correction.
ALFA is known to have significant first sidelobes, which have been mapped and are described in Heiles (2004). 
These sidelobes are displaced about $5^{\prime}$ from the main beams and have a typical sidelobe efficiency of 13\% for the central beam and 23\% for each of the six exterior beams. To correct for this effect, we use an algorithm similar to the one used in Hartmann et al. (1996\nocite{hartmann96}) for the Leiden-Dwingeloo Survey (LDS). We cannot use that algorithm in its exact form, as we sample the sky with 7 different beams rather than only one beam as in the LDS. In our modified method we determine the contamination to each of our observed spectra from the first sidelobe. We do this by determining the position each sidelobe has on the sky during each integration and calculating the spectrum the sidelobe would have added to the main beam data. We then subtract the `sidelobe contamination spectra' from original spectra, scaled by the beam efficiency, and use the `decontaminated' spectra to construct a new map.

The data we started with for examining M33 and its surroundings were gridded into a cube smoothed to 0.74 \kms~velocity resolution.  To bring out diffuse halo features we also created another cube smoothed to 5.15 \kms~resolution.  This latter cube was used for the majority of the analysis presented here and has a $3\sigma$ sensitivity of 0.1 K per channel, or $10^{18}$ cm$^{-2}$ in each $\sim5$ \kms~channel.  
The typical linewidth of halo clouds of the Milky Way is approximately 25 \kms~\citep{deheij}, indicating we are sensitive to these types of clouds with column densities above $5 \times 10^{18}$ cm$^{-2}$ (3$\sigma$).   The typical peak column density of Galactic HVCs is $\sim10^{19}$~\cm~ \citep{putman02}. 
The $3\sigma$ mass sensitivity to objects with linewidths of 25 \kms~at 730 kpc is $2.4 \times 10^4$ \Msun.  This mass is generally at least an order of magnitude lower than the Galactic HVC complexes with distance constraints (i.e., Thom et al. 2006, 2008; Wakker 2001\nocite{thom06,thom08,wakker01}) and the lowest HI mass Local Group dwarf galaxies \citep{grcevich09}.  Before making the figures that include the entire velocity range of M33, the emission from M33 was isolated in the south where it begins to overlap with Galactic emission (see Figure~\ref{hichans}).  Diffuse M33 features to the south of M33 that are indistiguishable from Galactic emission may be lost; however the extremely high velocity resolution of the GALFA-HI data make it the best data available for distinguishing M33 features from Galactic emission.

\section{Results}

\subsection{GALFA-HI Observations}

The integrated intensity HI map of M33 is shown in Figure~\ref{hi} with the column density contour at $3.2 \times 10^{20}$ cm$^{-2}$ labeled.   All of the star formation in M33 is within this contour according to the GALEX FUV map (see Figure~\ref{hifuv}).  Also labeled are a northern arc of gas that extends from the northern side of M33's disk and wraps back towards the disk and a southern cloud that extends in a filament back towards the galaxy.  The northern arc includes part of the known warp of M33 and extends out to projected distances of 11-18 kpc from the contour at $3.2 \times 10^{20}$ cm$^{-2}$ ($\sim 16-22$ kpc from the galaxy center), while the southern cloud and southern part of the warp (to the east of the cloud) extend $\sim$10-11 kpc from this contour of the disk.  Additional properties of these features are summarized in Table 1.  Other notable features evident in Figure~\ref{hi} are the large extent of the HI gas ($28 \times 38$ kpc at maximum) and three fingers of gas extending away from the extreme edges of the HI distribution to the east and west.   
The total HI mass of all of the emission shown is M$_{\rm HI} = 1.4 \times 10^9$~\Msun.   
   
Figure~\ref{hifuv} shows M33's HI column density contours over the far-UV (FUV; 1350-1750 \AA) GALEX image ($1\sigma$ surface brightness limit of 28.0 mag arcsec$^{-2}$; Thilker et al. 2005\nocite{thilker05}).   The FUV emission is largely limited to the inner $8 \times 15$ kpc.  The HI thus extends over 2-3 times as far as the stellar component traced by the UV light that is now commonly used to trace stellar clusters
forming in the outskirts of galaxies (Werk et al. 2010; Thilker et al. 2007\nocite{thilker07,werk10}).  An extension of M33's HI over the optical stellar component has also been noted (e.g., Corbelli et al. 1989\nocite{corbelli89}).  The FUV emission is confined within the HI column density contour at $3.2 \times 10^{20}$ cm$^{-2}$ labeled in Figure~\ref{hi}.   The compact nature of the contours near the disk make it difficult to identify which contour the star formation begins at, but it can most likely be taken in further to the contour at $5.9 \times 10^{20}$ cm$^{-2}$.   Since the gas beyond the $3.2 \times 10^{20}$ cm$^{-2}$ contour has a HI mass of $2.5 \times 10^8$ \Msun, over 18\% of the HI mass of M33 is beyond the star forming disk.

The complex velocity structure of M33's HI emission is best highlighted in the channel maps of Figure~\ref{hichans}. 
The channel maps have been extended from -412 to -46 \kms~(LSR)
to encompass a large high-velocity cloud known as Wright's cloud (Wright 1974, 1979\nocite{wright74,wright79}) and emission from the Galactic plane.   Each of the $\sim$5 \kms~channels in Figure~\ref{hichans}
shows gas with column densities as low as $10^{18}$ cm$^{-2}$. 
The intensity weighted velocity of M33's HI emission along a given sightline is also shown in Figure~\ref{hivel}.  All velocities are in the LSR reference frame.

The first feature to appear in the channel maps of Figure~\ref{hichans} is Wright's cloud.
Wright's cloud extends from the channels at -411 to -339 \kms~and spatially this cloud is known to continue
for at least another 1.5\deg~ to the west of the data shown here (Wright 1979; Lockman et al. 2002\nocite{lockman02}).    The structure of this entire complex as mapped by GALFA
will be presented in a future paper.
There is no evidence for a direct link between Wright's cloud and M33.   They are separated in 
velocity space by $\sim20$ \kms, and at this closest point in velocity they would be spatially separated by over 50 kpc at the distance of M33.  In the integrated intensity maps of both objects they appear to be separated by only $\sim 1.25$\deg\ (e.g. Figure~\ref{duchamp}), or 16 kpc at the distance of M33, but the velocity separation at this closest point is over 250 \kms.   The mass of the part of the complex shown here is 186 D$^2$(kpc) \Msun.

M33's HI emission begins at a velocity of -324 \kms~in Figure~\ref{hichans} and shows neither a diffuse link to Wright's cloud nor signs of a filament in the direction of M31.  
The emission subsequently extends both south into the main body of M33's disk 
and north into M33's known warp and the northern arc labeled in this figure and Figure~\ref{hi}.   
A finger of HI emission extends off of the northern arc at $\sim-241$ \kms~and the arc becomes basically indistinguishable from disk emission at about -200 \kms.    
The mass of the entire northern arc starting from where it emanates from the elliptical disk contours in Figure~\ref{hi} is $7.7\times10^{7}$ \Msun.

Continuing through the channel maps, much of the HI emission surrounding the high column density gas in M33's disk is difficult to classify into individual features.  This emission is diffuse in that it surrounds the majority of the disk and is at a relatively low column density compared to the disk emission.  In particular, this diffuse emission
surrounds the eastern side of the galaxy from about -200 to -100 \kms~and eventually merges with the southern warp. 
The southern cloud labeled in this figure and Figure~\ref{hi} is notable as a somewhat more isolated feature and is most evident in the channels from -159 to -138 \kms. 
This cloud links back to the disk of M33, and the filament from the cloud to the disk can no longer be easily isolated by the channel at -118 \kms.   This cloud has a HI mass of $1.4\times10^{6}$ \Msun\ at 730 kpc and was mapped at 
high resolution by \cite{westmeier05}.  There is another filament in the central regions north of the filament to the southern cloud that extends into more negative velocities and is obscured by the disk emission in the integrated maps.  
Galactic emission begins to overlap with M33's emission at velocities less negative than -107 \kms.  The last emission which is clearly part of M33 is at the tip of the southern warp at -46 \kms. 

Two other figures are included to represent additional properties of M33's HI emission.  The first is an intensity weighted
 velocity dispersion map of M33 (Figure~\ref{hiveldisp}).  The gas beyond the star forming disk generally has velocity dispersions $< 15$ \kms, while the gas in the star forming disk has a mean velocity dispersion of 18.5 \kms.  The central regions clearly have the highest
dispersions with values up to 30 km s$^{-1}$.   The decreasing values towards the outer regions are found for other local gas-rich galaxies \citep{tamburro09}.   Tamburro et al. assess the effect of beam smearing on the velocity dispersion of a galaxy (NGC 5055) with a slightly higher inclination than M33 (59\deg~ vs. M33's 56\deg) and a 265 pc beam, and find the contribution is at most 10\% in the outer regions and 20\% in the inner regions.
Finally, Figure~\ref{xray} shows a three dimensional map of M33's HI structure.   The disk is saturated to bring out M33's surrounding features.  The velocity range extends from -360 to -52 \kms, and therefore includes some Galactic emission to the right.  The northern arc is clear, as well as the southern warp and diffuse gas that connects back to the disk.

\subsection{Warped Disk Model Comparison}
\label{sec-warp}
We compare our data with the warped disk model of CS97 to examine how the features identified here are related to this model. The CS97 model is a series of nested rings, each with a HI surface density ($\Sigma\left(r\right)$), circular velocity ($V\left(r\right)$), inclination angle ($\phi\left(r\right)$), and major axis position angle ($\Theta\left(r\right)$). These parameters vary with ring radius using sigmoidal functions and a total of 14 independent parameters are used to specify these functions as outlined by CS97.  To directly compare this model with our HI data set we also required a line width parameter ($\sigma$) and a point-spread function (PSF). We use the standard Arecibo PSF of $3.4^{\prime}$ FWHM. To determine $\sigma$ we ran a least-squares fit of the model to the data allowing only $\sigma$ to vary, using the MPFIT.PRO program within IDL (Markwardt 2009)\nocite{markwardt09}. We find a best fit $\sigma$ = 14 km $s^{-1}$, very close to the value adopted in CS97, $\sigma$ = 12 km $s^{-1}$.

Comparisons between the warped disk model and the GALFA-HI data are shown in Figures~\ref{modchans} \& \ref{modvel}.  As noted by CS97, the warped disk (inclined $\sim$ 30\deg\ to the central disk) reproduces the twisted overall velocity structure of M33, but an additional component is needed to accurately reproduce M33's HI distribution in all areas.  The model over-predicts the HI flux in the northeast, and to a lesser extent in the south.  The most significant places where the model cannot reproduce the observed HI structure are in the majority of the features discussed above, including: the northern arc, the southern cloud, the filaments in the central regions (see Figure~\ref{modchans}), and the diffuse emission and fingers to the east and west.  There is also an offset in the velocity of the model relative to the data on the east and west sides of the galaxy.  For instance, in Figure~\ref{modvel} if one of the six central model contours is followed from the east to the west, the underlying observed HI emission is at a different velocity (color) on each side.  
This may represent the mixing in of the gas beyond this model radius.   The gas detected beyond the model radius tends to have more negative velocities than the gas detected within this radius to the west, and more positive velocities to the east.

\subsection{Cataloguing Discrete Features}
\label{sec-discrete}
Isolating halo gas from disk gas is difficult and can introduce false separations
of related features.   Besides the visual inspection of features described above, we
also approached the problem of defining discrete features by running an automated object finder
called Duchamp\footnote{http:$//$www.atnf.csiro.au$/$people$/$Matthew.Whiting$/$Duchamp$/$} on the data.  This allowed us to search for faint discrete objects that
may have been missed by eye and provided a quantitative method of distinguishing objects.   
Duchamp allows one to specify a detection threshold (initial seed of a detection must be above this level) and merges features that are linked in position and/or velocity at this threshold, or a different threshold if desired.
Duchamp therefore combines features that are linked by diffuse filaments and catalogs truly discrete features as individual objects.   

When running Duchamp we first applied a hanning smooth with a width of 5 channels to increase the chances of detecting low column density clouds with velocity widths similar to Galactic HVCs ($\sim$25 \kms).\footnote{We also ran Duchamp on the original velocity resolution cube, but this did not catalog additional discrete objects.}   We then ran several iterations of Duchamp and found using a detection threshold and merge parameter of 0.08 K (3$\sigma$ for this smoothed cube) was the most successful in terms of finding truly discrete clouds and not pushing too far into the noise.  The results of this run are shown in Figure~\ref{duchamp} and tabulated in Table 2.  Because of the increase in the noise level as Galactic emission is approached, we cut the search off at $-100$ \kms.  We therefore do not calculate the total flux of M33 (object \#11 in Figure~\ref{duchamp}) with this method and exclude two clouds detected at the edge of the velocity range that were clearly part of a Galactic filament (see Figure~\ref{hichans}).  We extended the search to $-450$ \kms~in the other velocity direction to include the clouds associated with Wright's cloud.  Each object catalogued by Duchamp was examined in the data cube to assess if it is a genuine detection.  This led to the removal of objects \# 8 and 10 from Table 2 and Figure~\ref{duchamp}.  

Independent of M33 and Galactic emission,   
12 discrete real objects were catalogued by Duchamp between -450 to -100 \kms~ in the region shown in Figure~\ref{duchamp} (Table 2).
7 objects have central LSR velocities between -400 to -350 \kms~ and are almost certainly associated with Wright's cloud (bottom half of Table 2).  Of these 7, objects \# 1, 3, and 7 are in a noisier part of the cube and as can be seen from Figure~\ref{hichans} could be considered part of object \#2, or the main part of Wright's cloud.  Object \#5 appears spatially removed from Wright's cloud here, but is most likely also associated, as an extension in this direction was also noted by Lockman et al. (2002) and Braun \& Thilker (2004)\nocite{lockman02,braun04}.  For the remaining 5 objects at the top of Table 2, \#9 is clearly related to the northern arc of M33 and \#12 to the southern cloud.  The other 3 objects (\#13, 14 \& 15) are at velocities between -130 to -113 \kms~ and are further from M33 (6-19 kpc from M33's HI edge in this figure at the distance of M33).  If located at the distance of M33 (730 kpc), the objects in the first part of Table 2 have an angular extent of 1-3 kpc, velocity widths of 17-30 \kms, and masses between $10^4$ and $2 \times 10^5$ \Msun.  

Though the data presented here only continue to declinations of +32.5\deg~at full sensitivity, we find no features north of what is shown in Figure~\ref{duchamp} beyond emission which is part of Galactic filaments.   The diffuse filament towards M31 claimed by Braun \& Thilker (2004\nocite{braun04}) is not continuous in velocity with M33 and has LSR velocities generally less negative than -100 \kms.  This detection may be part of the diffuse filaments extending into our Galaxy partially shown in the channel maps of Figure~\ref{hichans}.  The heavy kinematic and spatial smoothing used in Braun \& Thilker (2004) may
have blended parts of the Galactic emission into their maps.   Future deep mapping of the region between M31 and M33 will resolve the origin of any clouds found in this region.

\section{Orbital History of M33}
\label{sec-orbit}
To explore the origins of M33's gaseous features we
investigated the feasibility of a tidal interaction between M33
and M31 by constraining the orbital history of the system.  
Given the large present
separation between M33 and M31, a significant tidal interaction that could lead
to the gaseous features seen here is
only possible if past orbits brought the two galaxies closer to each
other.  This modeling also puts constraints on the presence of satellites around M33, as they would not remain bound to M33 if it had a previous close encounter with the more massive M31.

The orbital history of M33 can be constrained by integrating the motion of M33 backwards in time through
M31's evolving gravitational potential.  Since the proper motion of
M33 has been recently measured ($190 \pm 59$ \kms\ relative to Earth;
Brunthaler et al. 2005\nocite{brunthaler05}) and the radial velocities
of both galaxies are well known ($-39$ \kms\ for M33 and $-116$ \kms\
for M31 relative to the Galactic center), the only unknown velocity
components are the tangential velocities of M31 in the observed
reference frame.  
We calculated a large set of possible orbits for the two
tangential velocity components of M31, ranging each from $-200$ to
$+200$ \kms.  Higher velocities are unlikely in a relatively poor group of galaxies such as the Local Group.  Our
grid selection method is similar to that in Loeb et al. (2005).  We
treat these possible orbits as a statistical ensemble and use it to
estimate the probability of a recent close encounter.  Following van der Marel \& Guhathakurta (2008), we use a distance of 770 kpc for M31 and 794 kpc for M33.  The variation in the distance for M33, compared to the 730 kpc from Brunthaler et al. (2005), does not affect our results within the uncertainties.

For the purpose of orbit calculations, we treat M33 as a point mass
moving in the combined gravitational potential of M31 and the Galaxy.
The mass distribution of M31 is modeled as the sum of a baryon disk
and a dark matter halo, with the total virial mass $M_{\rm vir}(0) =
2\times 10^{12}$ \Msun\ at redshift $z=0$.  We adopt the disk mass $M_d(0) =
4\times 10^{10}$ \Msun.  The disk is described by an analytical
Miyamoto-Nagai profile with a scale length $r_d = 5$ kpc.  Although
our adopted baryon mass may be an underestimate of the M31 disk and bulge,
we show below that it plays a much smaller role than the dark halo in
determining the possible orbits of M33.  A more detailed model of the
M31 stellar distribution by Fardal et al. (2007\nocite{fardal07}) had a similar scale length but larger
total baryon mass.

For the halo we adopt a NFW profile with a concentration parameter
$c=12$ at $z=0$, in line with the expectations of $\Lambda$CDM
simulations.  This results in a scale radius $r_s \approx 25$ kpc.
The halo mass distribution is the main uncertainty of our
calculations.  Unfortunately, the outer mass distribution of M31 is
not well determined observationally.  Fardal et al. (2007) and Seigar
et al. (2008\nocite{seigar08})
calculate $M_{\rm vir} \approx 10^{12}$ \Msun, while Loeb et
al. (2005) and van der Marel \& Guhathakurta (2008) adopt $M_{\rm vir}
\approx 3 \times 10^{12}$ \Msun.  Such widely different estimates can 
be partly due to the different analysis and partly due to the
different assumed profiles.  In particular, Fardal et al. (2007) find
the best fit to the rotation curve requires the concentration
parameter $c \sim 30$, much higher than that expected for $\Lambda$CDM
halos of such mass.  A very concentrated lower-mass halo and a less
concentrated higher-mass halo can contain similar amounts of mass
within a given radius.  For example, in our model the total masses of
stars and dark matter enclosed within 60, 120, and 200 kpc are $(0.66,
1.15, 1.60) \times 10^{12}$ \Msun, respectively, while in the Fardal et
al. (2007) model these masses are $(0.51, 0.72, 0.88) \times 10^{12}$
\Msun.  
As we find no plausible
orbits of M33 coming closer to M31 than 60 kpc, the size of the baryon
distribution in unimportant for this calculation.

Since the mass of a galaxy is expected to grow with time due to
hierarchical merging and accretion, we account for this evolution by
reducing the mass as we integrate the orbits in the past for 10 Gyr
from the present time, $t_0$.  We assume that the disk mass scales
linearly with cosmic time, $M_d(t) = M_d(0) \, t/t_0$.  For the halo
mass, we adopt an exponential growth \citep{wechsler02} with the
formation redshift $z \approx 1$, which gives $M_{\rm vir}(z) = M_{\rm
vir}(0) \, e^{-z}$.  Note that for most orbits we find the last close passage between M33 and M31 within the last 3 Gyr ($z < 0.3$).  In this relatively short interval the gravitational potential is not expected to have changed significantly, and therefore, our assumptions about its evolution with redshift are not affecting our conclusions.

In addition to M31, we include the gravitational potential of the Galaxy,
treating it as a point-mass ($10^{12}$ \Msun) with an analogous
redshift evolution.  It turns out to have a negligible effect due to the
large current distance to M33 and an even larger separation in
the past.  We also include dynamical friction on the orbit of M33
\citep{binney08}, but the effect is again small because the galaxy
never gets close enough to M31.

The main goal of the orbit calculations was to determine the evolution
of the tidal radius of M33, $r_t$.
Any material outside $r_t$ is no longer gravitationally bound to the
small galaxy and would be pulled away by tidal forces.  However, it
should be noted that the isopotential surfaces are not spherical and a
single value of the tidal radius gives only a rough idea of the tidal
truncation of M33.  To estimate $r_t$ we assume flat
rotation curves for both M33 and M31 around $r_t$, with the circular
velocities $V_{M33} \approx 100$ \kms\ and $V_{M31} \approx 250$ \kms,
respectively.  This gives $r_t \approx 2/3^{3/2} \, R_p \,
(V_{M33}/V_{M31}) \approx 0.15 \, R_p$, where $R_{p}$ is the
perigalactic distance between M33 and M31.

The results of the orbit calculations are shown in
Figures~\ref{fig:sim1}--\ref{fig:veltime}.  Figure~\ref{fig:sim1}
shows the probability (cumulative fraction of orbits) of a given tidal
radius for the orbits with the maximum tangential velocity of
M31 restricted to be below 50, 100, and 200 \kms.
Although about 30\% of the simulated orbits do not have a previous
closer approach, these orbits seem unlikely as it would imply that M33
did not form in the vicinity of the other galaxies of the Local Group.
Figure~\ref{fig:sim1} shows that at all
possible M31 velocities leading to a previous closer approach between
M33 and M31, there is a 60\% probability of the galaxies once being
within 100 kpc of each other, resulting in a tidal radius of M33 below
15 kpc.  For the most likely scenario of V(M31)$_{tan} < 150$ \kms~
(see also \cite{vandermarel08} who find V(M31)$_{tan} < 56.3$ \kms),
there is more than a 75\% chance of the M33 tidal radius being smaller than
15 kpc.  Even if we include the orbits lacking a previous closer
approach, at these most likely M31 velocities the probability for $r_t < 15$ kpc is
still very high, about 60\%.
Figure~\ref{fig:m33orbit} shows a 3D representation of one of such
orbits with $r_t \approx 10$ kpc.

Figure~\ref{fig:radtime} shows the probability distribution of
the distance and lookback time of the close encounters for the entire
range of velocities.  The most likely
encounters are the closest and most recent (darker shading on the
plot).  Approximately a third of all possible 
orbits lead to $R_p < 80$ kpc and
$r_t < 12$ kpc, which occurred between 1 and 2.5 Gyr ago, and half result in a
close encounter resulting in $r_t < 15$ kpc within the last
3 Gyr.  Even though this is a statistical argument, it provides
support for the idea of a relatively recent tidal interaction between
M33 and M31 even if the M31 velocity is much higher than expected.  
Figure~\ref{fig:veltime} shows the probability distribution of the M33
velocity relative to M31 at the last close encounter.  The majority of
the orbits indicate a velocity around 400 \kms.

\section{Discussion}

M33's gaseous features are connected to the galaxy in position and velocity and cause M33 to extend over $28 \times 38$ kpc at maximum extent.  Given the likelihood of a previous closer approach with M31 resulting in a tidal radius $<15$ kpc (see Section 4), these features were most likely pulled from the outer disk of M33 during this passage.  The gaseous disks of
spiral and dwarf galaxies are commonly found to extend significantly further than the stellar components \citep{meurer96, begum05} and will be the first thing stripped from a galaxy during an interaction.   The early discovery
of M33's warp was an initial hint of M33's disruption (Rogstad, Wright \& Lockhart 1976; Corbelli \& Schneider 1997\nocite{rogstad76, corbelli97}), but at that time it was concluded that the tidal force of
M31 relative to M33 was not strong enough to cause the warp.   
The recent proper motion measurements of M33 have permitted the orbit of this galaxy relative to M31 to be assessed.  Our analysis, as well as the analysis of Loeb et al. (2005\nocite{loeb05}) and van der Marel \& Guhathakurta (2008\nocite{vandermarel08}), indicate a previous interaction between M33 and M31 is highly likely.    The tidal radius of $<15$ kpc and approach of $<100$ kpc within the past 3 Gyr occurs in more than 75\% of the most likely orbits (V$_{M31} < 150$ \kms) in our analysis.  Given M33's gaseous features extend out beyond this radius and the largest features are relatively symmetric, a tidal origin seems likely.  In addition, since a satellite would not remain bound to M33 unless located at $< 15$ kpc, the gaseous features are unlikely to be the remnants
of an accreting satellite.   Deep stellar surveys have also not revealed any clearly distinct stellar features in the vicinity of M33 (e.g., Ibata et al. 2007\nocite{ibata07}).

As M33 passed through M31's extended halo at most likely distances of 65-120 kpc (Figure~\ref{fig:radtime}), ram pressure forces from M31's hot gaseous halo are also likely to have acted on M33's gas.   Though there is no direct evidence for a hot halo around M31, it is inferred to exist through the lack of gas in the dwarf satellites of M31 \citep{grcevich09}, head-tail halo clouds \citep{westmeier05}, and the predictions of simulations of galaxy formation \citep{sommer06, kaufmann06}.    Using the simple Gunn \& Gott criterion \citep{gunn72} as used by Vollmer et al. (2008)\nocite{vollmer08},
\begin{equation}
\rho_{IGM} \sim \frac{v_{rot}^{2}\,\Sigma_{gas}}{v_{gal}^{2}\, R} \; cm^{-3},
\end{equation}
we can estimate if M31's hot halo is likely to have played a role in stripping the outer parts of M33.  M33 would be moving at approximately $v_{gal} = 300-415$ \kms~relative to M31 at perigalacticon (Figure~\ref{fig:veltime}). Using a rotational velocity of $v_{rot}=100$ \kms~\citep{ciardullo04, corbellisalucci00}, a stripping radius (R) of 9 kpc (lower bound corresponding to the edge of the FUV disk), and a $\Sigma_{gas}$ of $3 \times 10^{20}$ cm$^{-2}$ (see Table 1)
suggests M31 halo densities greater than $10^{-3.5}$ cm$^{-3}$ are needed for ram pressure stripping to be effective on the outer regions of M33.  
 At distances of 65-120 kpc a hot halo around galaxies similar to M31 is highly unlikely to have these types of densities \citep{bregman07, peek07, gaensler08}.  Densities on the order of $10^{-4.5} - 10^{-3.5}$ cm$^{-3}$ are expected from the galaxy formation simulations at these radii \citep{sommer06, kaufmann06} and the observational constraints referenced above are most consistent with the lowest end of this range.  Therefore, the role of a diffuse halo medium around M31 in stripping out M33's gaseous features is likely to be minor, consistent with their symmetric structure not being indicative of ram pressure stripping \citep{vollmer08, chung07}.

Though ram pressure effects are unlikely to lead to the origin of M33's gaseous features, they will apply a drag on this relatively low column density gas (compared to the disk) at large radii from the center of M33's gravitational potential.   The fingers extending from the sides of the galaxy may be evidence of this effect.  The combination of this drag and the tidal forces will cause M33's outer gas to lose angular momentum and fall towards the central regions of M33 as it moves away from the gravitational influence of M31 \citep{moore94, gardiner99, barnes02, cox08}.  None of the HI gas surrounding M33's stellar disk is above M33's escape velocity and its infall may be subtly evident in its kinematics.
The gas on the western side of M33 has more negative velocities than the gas in the underlying disk, consistent with an infall of gas from the far side of the galaxy; while the gas on the eastern side has more positive velocities than the gas in the disk consistent with a net inflow from the near side of the galaxy.  
 Gas falling back onto M33 may represent an important method of fueling and triggering the star formation in M33's disk (see Section~5.2.2).

\subsection{M33 Halo Clouds}

The limited number of discrete objects catalogued by Duchamp in the vicinity of M33 (Section 3.1) highlights that the majority of the surrounding gas is directly connected to M33's disk in position-velocity space.  This includes the majority of the clouds catalogued by \cite{grossi08}.  The clouds that are catalogued as discrete objects (see Figure~\ref{duchamp}) are either very close to being connected to M33 in position-velocity space (\#9, 12), associated with the Wright cloud complex, or are the three clouds at low velocities (V$_{\rm GSR}=4-20$ \kms; \#13, 14, 15).   These lower velocity clouds may be part of a Galactic filament or they may be related to M33, as they do extend towards the position and velocity of a filament on that side of the galaxy.  If related they extend $\sim$19 kpc from the HI edge of M33 ($\sim$38 kpc from the center) and may represent additional material disrupted during the interaction with M31, or possibly some type of cold flow of gas towards the disk \citep{keres05}.  Deeper observations will help to confirm their relationship, if any, to M33.

The properties of the clouds that are associated (\#9, 12) or potentially associated (\#13,14,15) with M33 (top part of Table 2) can be compared to the halo clouds of our Galaxy and M31 (referred to as high-velocity clouds (HVCs) for our Galaxy).
The clouds have similar properties to HVCs and M31 halo clouds in terms of their column densities and velocity widths \citep{putman02, deheij, westmeier05}.   
The similar velocity width of the clouds (20-30 \kms) is indicative of a warm cloud component ($\sim$10,000 K) and suggests a similar nature for the M33, M31 and Milky Way halo clouds \citep{stanimirovic08, bekhti06, westmeier05}.
For Galactic HVCs this warm component is often interpreted as representative of a pressure equilibrium with a surrounding hot halo medium \citep{wolfire95}.   M33 is not generally expected to have a remnant primordial hot halo as it is not considered massive enough for a large quantity of gas to be shock-heated as it flows into the dark matter potential well \citep{keres05, brooks09}.  In addition, if there was any surrounding halo medium it would most likely have been stripped during the encounter with M31.   Therefore the warm cloud component may represent warm material stripped from M33, or heating from either the interaction or escaping radiation from M33's disk \citep{blandhawthorn99}.  Sensitive Fabry-Perot H$\alpha$ observations of the extended gaseous M33 system would be a very interesting complement to the data presented here.

At the distance of M33, the discrete clouds also have similar masses to the M31 clouds and the large
Galactic HVCs with distance constraints ($\sim$10 kpc; \cite{thom06, thom08, wakker08}); but the physical sizes tend to be much smaller.  This may be consistent with the largest clouds catalogued as halo clouds in M31 and the Galaxy either showing a connection to the disk, or in the case of the Galaxy, the fact that they would show a connection if observed by someone in an external galaxy  \citep{thilker04, kalberla05}.  For instance, for M33 the southern cloud is connected to the gaseous disk (see Figure~\ref{hi}), and is more typical in size (though still on the small side) to Galactic HVCs.  Its mass would be similar to some of the larger Galactic HVC complexes.  In any case, the clouds surrounding M33 show numerous similar properties to the Galactic and M31 halo clouds and this could be suggestive of a similar origin or destruction mechanism. 

The clouds catalogued as part of the Wright's cloud complex are unlikely to be associated with M33 given their projected distance from the galaxy, velocity offset, and the large mass and size of the complex at the distance of M33 ($>10^8$ \Msun; \cite{wright79}).  This cloud may be associated with the tail of the Magellanic Stream, as the Stream is found to extend quite far in the direction of M33 and has several features at similar extreme negative LSR velocities \citep{stanimirovic08,westmeier08}.  At 60 kpc this cloud would have a mass on the order of $10^6$ \Msun.  The linewidths of the catalogued clouds that are part of the Wright's cloud complex are again typical of Galactic HVCs, as is the angular size of the primary part of the complex.    
The GALFA data of this complex will be analyzed further in a future paper with a similar method to that outlined in Peek et al. (2007)\nocite{peek07}.

\subsection{Future of M33}
\label{sec-fuel}
\subsubsection{Fueling M31}

The most likely fate of M33 is to fuel M31 given the current disruption of this galaxy and their likely motion relative to each other.   M33 will therefore provide $1.4 \times 10^9$ \Msun\footnote{The molecular component of M33 is estimated to be 
only an additional $4.5 \times 10^7$ \Msun\ \citep{engargiola03}.}  in star formation fuel to M31, which currently has only $5-6 \times 10^9$ \Msun\ in fuel.   
Depending on the orbits of the galaxies, M33 may eventually resemble the Magellanic System, with leading and trailing streams of gas being stripped
through some combination of tidal and ram pressure forces \citep{putman03, connors06b, mastropietro05}.  It is possible that some of the clouds in M31's halo were left behind from the first
passage of M33.   In particular there are several clouds at projected distances up to 50 kpc from M31's disk that
cannot be associated with any known stellar features \citep{westmeier07, thilker04}.  

Mergers with large satellites may be an important source of star formation fuel for spiral galaxies like M31 and the Milky Way.  The Magellanic System and the Milky Way are even closer in total gas mass than M33 and M31, with the Magellanic System hosting $\sim 10^9$ \Msun, close to 50\% of the Milky Way's gas mass \citep{putman03, levine06}.   Chemical evolution models for the Galaxy typically require an average inflow of $\sim1$ \Msun$/$yr \citep{chiappini01, fenner03}, and obtaining
all of this from lower mass satellites, or gas condensing from the hot halo appears to
be difficult \citep{peek08, putman06, grcevich09}.  The equivalent of one system
of similar gas mass to the Magellanic System or M33 per Gyr over the past 5-7 Gyr would have provided a galaxy like the Milky Way or M31 with enough fuel to maintain its
star formation and not become severely depleted.

A scenario of larger, gas-rich, satellite mergers providing the majority of the star formation fuel
for a spiral galaxy may also be consistent with the survival of a galaxy disk according to recent simulations \citep{robertson06, brook04, moster09}.    Though the M31:M33 total mass ratio is not quite at the 1:10 mass ratio often considered particularly harmful to the disk, at approximately 1:20-40 it is significantly larger than the majority of the dwarf satellites generally considered as fuel for Local Group galaxies.   The simulations generally find that when the merging system is gas dominated, the thin disk is either able to re-form as the gas settles back onto the disk and/or the kinetic impact energy is absorbed by the gas. 
The evolution of satellites towards radial orbits as they merge may also substantially suppress the disk heating \citep{hopkins08}, although the addition of gas seems to be a more widely accepted result in aiding disk survival \citep{purcell09}. 
The presence of M33 and the Magellanic System in the context of M31 and the Milky Way's evolution indicates these types of gas-rich mergers continue to occur at $z=0$.

\subsubsection{Fueling M33}

M33 is currently forming stars at $\sim0.7$ \Msun$/$yr \citep{blitz06}.  The galaxy's HI mass indicates it only has 2 Gyr left of star formation activity at this rate unless a source of fuel is obtained from the Local Group.  Thus M33 appears to be in a similar situation to the Milky Way and other local spiral galaxies (e.g., Peek 2009\nocite{peek09}); it is either in a unique place in time where it will soon run out of star formation fuel, or it has a continuous mode of obtaining star formation fuel from a surrounding medium.  Recent chemical evolution models of M33 also suggest continuous gaseous inflow is needed over the lifetime of the galaxy to explain the range of stellar metallicities \citep{magrini07, barker08}.

A source of continuous fuel for M33 is difficult to find.
M33 may be partially fueling itself with the gaseous features that extend beyond the current star forming disk, but this is only $2.5\times10^{8}$ \Msun\ and would rapidly be exhausted.  Satellites cannot provide M33 with fuel given the close passage with M31 would have stripped them from M33.  As mentioned previously, the lack of satellite accretion for M33 is also consistent with its smooth stellar distribution in deep surveys to date \citep{ferguson07}.   The replenishment of M33's HI through cooling of gas near the disk is a possibility given the connection of M33's gaseous features to the star forming disk.  The Milky Way, M31, and other spiral galaxies also tend to have the majority of their halo gas close to the disk, and the higher densities there make cooling of a diffuse halo medium to HI a more likely possibility \citep{heitsch09, peek08, fraternali08}.
If this is the case, the source of the cooling gas remains a mystery, as the passage with M31 is likely to have stripped any M33 halo medium and there is no strong evidence for a substantial diffuse Local Group medium.

Despite the limited mass present in M33's gaseous features for long term feeding, on the short term the inflow of this gas back towards M33 may be affecting its star formation.
The pressure of the infalling gas may help to explain the efficiency of M33's current
star formation.  Several groups have found that M33 is more efficient at converting molecular gas to stars than typical nearby spirals \citep{gardan07,blitz06,kennicutt98}.
For instance, \cite{blitz06} are able to predict the molecular gas fraction with their pressure regulated star formation prescription, but under-predict the star formation rate by a factor of 7.   An additional pressure on the gas
generated by material falling back onto the disk and the higher velocity dispersion in the disk than assumed for all galaxies
by \cite{blitz06} may be a component in the discrepancy.    Blitz \& Rosolowsky assumed a constant velocity dispersion of 8 \kms~across
the disks of all of their galaxies, and their mid-plane pressure is directly proportional to this value.   Still M33's higher mean value of 18.5 \kms\  (see Figure~\ref{hiveldisp}) should affect the molecular gas
fraction as well as the SFR under their criterion, so this is not a simple solution to the discrepancy in SFR.   We note that M33's mean velocity dispersion is also elevated compared to the sample of gas-rich galaxies studied by \cite{tamburro09}, with the exception of NGC 5194, which they argue is elevated due to the enhanced star formation triggered by the tidal interaction with NGC 5195.  Since M33 is now beyond the dominant gravitational influence of M31, perhaps it will soon cycle out of an elevated star formation phase.

M33 does not appear to show star formation in its surrounding gaseous structures, or an extended UV (XUV) disk as seen for many galaxies with large gaseous features beyond the main optical component (see Figure~\ref{hifuv}; \cite{werk10,thilker07}).  
We compared our observed column densities at the largest radii M33 is forming stars to the expected critical surface density derived by Kennicutt (1998) based on the Toomre (1964\nocite{toomre64}) criterion and confirmed previous findings that the observed HI column (3 - 5 \Msun~pc$^{-2}$) is three times lower than this expected critical density.  Therefore
M33 is forming stars in relatively low column density gas, but for some reason this star formation does not continue into the extended gaseous features like it does in other systems.  The presence (or lack of) star formation in extended gaseous features will be an ongoing area of study, with M33 being an important local probe of this process.

\section{Conclusions}

\begin{figure}
\vspace{0.3in}
\epsscale{0.9}
\plotone{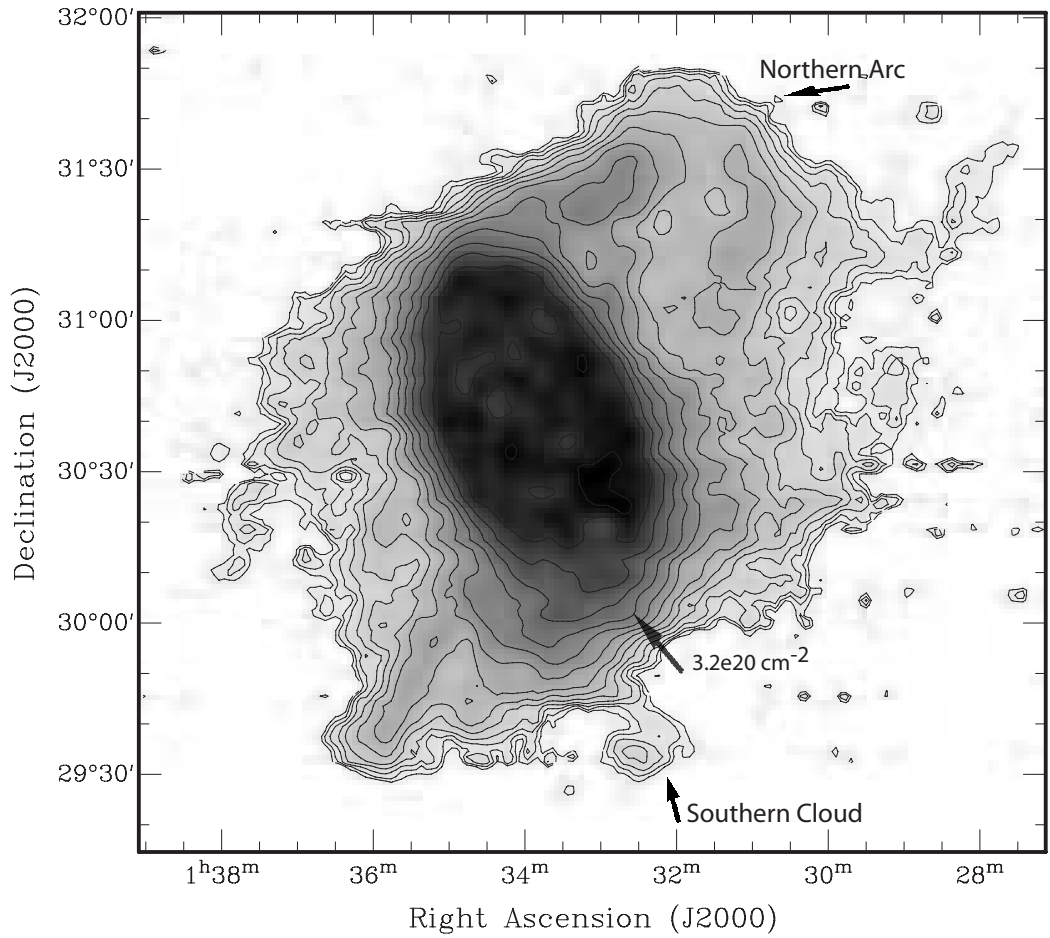}
\caption{HI column density map of M33.  Contours are $8.3 \times 1.5^n \times 10^{18}$ cm$^{-2}$ with n=0...13 and 8.3 $\times 10^{18}$ cm$^{-2}$ being 5$\sigma$ to a 25 \kms~feature. The maximum contour is at $1.6 \times 10^{21}$ cm$^{-2}$.  The beam is 3.4\arcmin~which is 720 pc at the distance of M33. 
}  \label{hi}
\end{figure}

\begin{figure}
\vspace{0.1in}
\includegraphics[angle=0,scale=0.8]{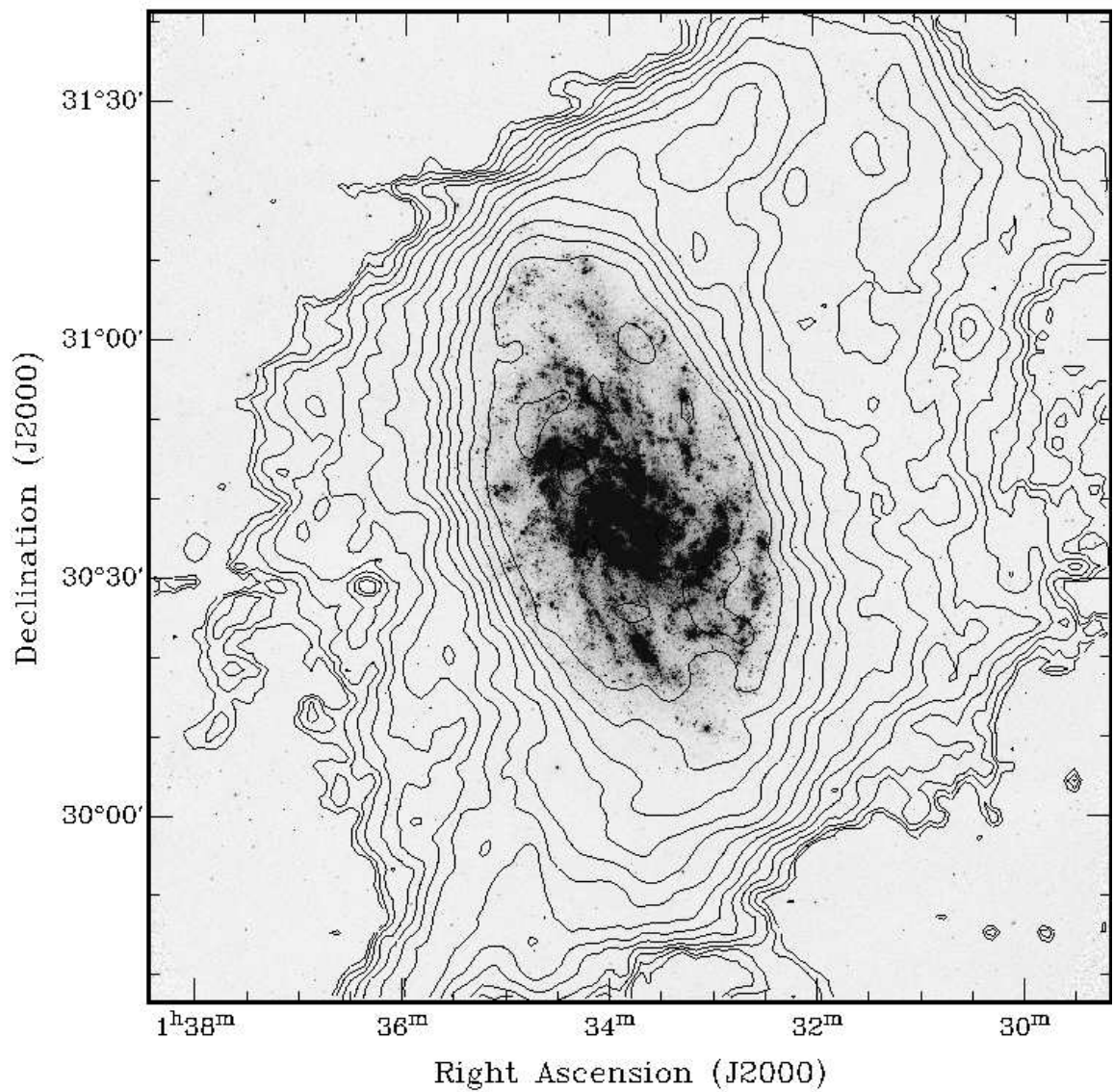}
\epsscale{1.3}
\caption{The Far-UV GALEX image of M33 with contours representing HI column density overlaid.  The contours are the same as Figure~\ref{hi}.  The FUV emission is largely confined within the $5.9 \times 10^{20}$ cm$^{-2}$ contour (the 12th contour counting from the outside).  }
\label{hifuv}
\end{figure}

We have presented new HI data of M33 and its surroundings from the GALFA-HI Survey and have highlighted several distinct gaseous features that extend out to 18 kpc from the edge of the stellar component traced in FUV emission (Figure~\ref{hi} \& \ref{hifuv}; Section 3.1).  The maximum extent of M33's HI is 28 $\times$ 38 kpc, and includes features such as a northern arc of gas that extends from M33's warp back towards the disk, a southern cloud with a filament to the galaxy, and diffuse emission and several fingers to the east and west of the galaxy (Figures~1-6).   The HI gas beyond the star-forming disk contains 18\% of M33's total HI mass of $1.4 \times 10^9$ \Msun\ and the vast majority of the gas surrounding M33 is directly connected to the disk in position-velocity space.  The GALFA-HI data is compared to the warped disk model of CS97 to highlight the relationship of the extended gas to this model (Figures~\ref{modchans} \& \ref{modvel}; Section 3.2). 
The following additional results are described in the paper.
\begin{itemize}

\item
The orbit of M33 relative to M31 is examined to investigate the feasibility of a tidal interaction between the two galaxies resulting in M33's gaseous features (Section~\ref{sec-orbit}).  Consistent with previous analyses of the likely orbits, we find a previous close encounter ($<$100 kpc) between the two galaxies is highly likely ($> 75\%$ of the orbits for the most likely V(M31)$_{tan} < 150$ \kms), resulting in a tidal radius for M33 of $< 15$ kpc.   This suggests a satellite of M33 would not remain bound to it and the gaseous features were most likely tidally pulled directly from M33 in the past 3 Gyr.

\item
The combination of tidal and ram pressure forces acting on M33 as it passes through an extended halo of M31, is likely to cause the tidally extracted gas to lose angular momentum and fall back towards M33's disk as it moves away from M31.   This is consistent with the velocity sense of the observed gas and the elevated star formation in M33's disk.

\item
An automated object finder is run on the M33 cube to define truly discrete objects from those directly connected to the galaxy (Sections 3.3 \& 5.1).  Besides the clouds associated with the Wright's cloud complex (which is unlikely to be associated with M33 given its offset and large size at M33's distance), there are only 5 small clouds catalogued, and all other gas is connected to M33.  Of the 5 clouds, 2 are very close to being connected to M33, and the other 3 are either in a filament going out 38 kpc from the center of M33 or part of a Galactic filament given their lower velocities.  These clouds show similar properties to Galactic HVCs and M31 halo clouds, including their velocity widths ($\sim25$ \kms) which are indicative of a warm cloud component.

\item
M33 will eventually become star formation fuel for M31, and if it does not substantially deplete its gas content before then, it will provide $\sim$25\% of M31's current HI mass.   This together with the Magellanic System, which hosts close to 50\% of the Milky Way's current HI mass, indicates the contribution of gas from large mergers can play an important role in the ongoing fueling of galaxies.  The fact that satellite mergers that are gas-rich are not expected to be overly disruptive to the disk is also support for these types of mergers providing substantial fuel for a spiral galaxy. 

\item
M33 will either run out of star formation fuel in the next 2 Gyr if it continues to form stars at its current rate, or it must accrete gas from an unknown Local Group medium.  Other notable characteristics of M33's gas and star formation are the high mean velocity dispersion across the disk ($\sim18.5$ \kms; Figure~\ref{hiveldisp}) that may be consistent with the relatively high SFR and recent interaction, and the fact that 
M33 does not have star formation as traced in the FUV in its extended gaseous component, or beyond the contours at $3-6 \times 10^{20}$ cm$^{-2}$ (Figure~\ref{hifuv}).

\end{itemize}

\acknowledgments
We acknowledge the ALFALFA team \citep{giovanelli05} and the wonderful staff at the Arecibo Observatory for their help in obtaining the data presented here, and
useful comments from James Bullock, Shika, and the referee.
MEP, SS, CH, EJK, and JEGP acknowledge support from NSF grants AST-0707597, 0917810, 0707679 and 0709347.
OG and AM are supported in part by NSF grant AST-0708087.  KAD received funding from the European Community's
Seventh Framework Programme under grant agreement n$^o$ PIIF-GA-2008-221289.
MEP also acknowledges support from the Research Corporation.
We credit the use of the Karma visualization software \citep{gooch96}.  This research made use of the Duchamp source finder, produced at
the Australia Telescope National Facility, CSIRO, by M. Whiting.
The Arecibo Observatory is part of the National Astronomy and Ionosphere Center, which is operated by Cornell University under a cooperative agreement with the National Science Foundation.

\bibliography{ref}
\bibliographystyle{apj}

\begin{deluxetable}{cccccc}
  \tabletypesize{\small}
  \tablewidth{0pt}
  \tablecaption{Properties of M33's HI}\label{tab1}
  \tablehead{\colhead{Region}& \colhead{RA\tablenotemark{a}}& \colhead{Dec\tablenotemark{a}}&\colhead{$V_{\rm LSR}$\tablenotemark{a}}&
                        \colhead{HI Mass \tablenotemark{b}}&\colhead{N$_{\rm HI}$ (peak)}\\
   		 \colhead{}&\colhead{}&\colhead{}&\colhead{km s$^{-1}$}& \colhead{\Msun}&\colhead{cm$^{-2}$} }
  \startdata
  
   Entire Galaxy &  01:33:42    &  30:35:30 &  -147  &$1.4\times10^{9}$ & $1.9\times10^{21}$ \\
   Gas Beyond Star Forming Disk\tablenotemark{c} &   -   &  -  &  -  &  $2.5\times10^{8}$ & $3.1\times10^{20} $\\
  Northern Arc\tablenotemark{d} & 01:31:50 &  31:18:30 & -232  &  $7.7\times10^{7}$ & $3.1\times10^{20} $ \\
  Southern Cloud\tablenotemark{e} & 01:32:30 &  29:35:30 &  -158 & $1.4\times10^{6}$ & $3.7\times10^{19}$ 
  \enddata
   \tablenotetext{a} {\scriptsize Approximate central value.}
  \tablenotetext{b} {\scriptsize At the distance of 730 kpc.  The dependency is M$_{\rm HI}~ \alpha$ D$^2$, so if the largest M33 distance of 964 kpc \citep{bonanos06} is adopted all HI masses should be multiplied by 1.7.}
  \tablenotetext{c}{\scriptsize Gas beyond the $\sim3.2\times10^{20}$ cm$^{-2}$ contour in Fig.~\ref{hi}.}
  \tablenotetext{d}{\scriptsize The full arc as it emanates from the regular disk structure in Fig.~\ref{hi}.}
  \tablenotetext{e}{\scriptsize Only the discrete part of the cloud evident in Fig.~\ref{hi}, i.e., does not include the filament back to the central regions of the disk evident in Fig.~\ref{hichans}.}
\end{deluxetable}

\clearpage
\begin{deluxetable}{lcccccccccccc}
\tabletypesize{\scriptsize}
  \tablewidth{0pt}
  \tablecaption{Discrete HI Sources at V$_{LSR}<-100$ \kms~ in Figure~\ref{duchamp}}
 \tablehead{
 \colhead{Obj \#} &    
 \colhead{RA} &      
 \colhead{DEC} &      
  \colhead{w$_{RA}$} &      
  \colhead{w$_{DEC}$} & 
  \colhead{$\Delta$RA\tablenotemark{a}} & 
  \colhead{$\Delta$Dec\tablenotemark{a}} &   
  \colhead{T$_{peak}$} &   
  \colhead{V$_{LSR}$} &   
  \colhead{V$_{GSR}$} &   
  \colhead{w$_{vel}$} &    
  \colhead{$\int T \Delta v$} & 
  \colhead{M$_{HI}$\tablenotemark{a}} \\
   \colhead{}            &       
   \colhead{}      &       
   \colhead{}      &    
   \colhead{$\prime$} &    
   \colhead{$\prime$} &      
   \colhead{kpc} &    
   \colhead{kpc} &     
   \colhead{K} &      
   \colhead{\kms} &         
   \colhead{\kms} &      
   \colhead{\kms} &       
   \colhead{K \kms}      &       
   \colhead{M$_{\odot}$} } 
\startdata  

       9 & 01:28:42.8 & +31:41:23.5 &       7.7 &        7.0 & 
             1.63 &       1.49 &      0.25 &     -264 &  -123 &    30.0 &   237 &    1.6E+05 \\
         12 & 01:33:25.3 & +29:26:50.1 &        6.1 &        6.0 &     
           1.30 &       1.27 &      0.22 &     -150 & -16  &    29.7 &   148 &      9.7E+04 \\
         13 & 01:39:33.1 & +30:41:21.0 &       12.0 &        8.0 &   
             2.56 &       1.70 &      0.20 &     -129 &  4 &     28.5 &   202 &    1.3E+05 \\
         14 & 01:44:01.0 & +31:15:52.4 &        9.4 &        5.0 &     
           2.00 &       1.06 &      0.22 &     -127 &   5 &   20.9 &   92.5 &     6.1E+04 \\
         15 & 01:40:08.1 & +30:49:17.4 &        5.2 &        5.0 &      
          1.09 &       1.06 &      0.24 &     -113 &    20 &  17.3 &   55.7 &    3.7E+04 \\
\hline
Wright & Cloud & Objects & & & & & & & \\
\hline
   1 & 01:15:18.5 & +30:49:26.8 &       17.2 &       11.0 &  
       3.65 &       2.34 &      0.19 &     -391 &   -246 &   37.6 &   236 &     1.6E+05 \\
 2 & 01:20:21.5 & +29:42:17.1 &      182.4 &       92.0 &    
       38.73 &      19.54 &      4.33 &     -390  &  -249 &    17.7 &   1.35E+05 &     8.9E+07 \\
3\tablenotemark{b} & 01:14:17.6 & +30:40:46.4 &        9.5 &        6.0 &    
         2.01 &       1.27 &      0.17 &     -383 &  -238 &     16.1 &   29.8 &      2.0E+04 \\
4 & 01:19:02.0 & +31:13:01.5 &       13.7 &       13.0 &     
      2.90 &       2.76 &      0.39 &     -380 &  -236  &  24.0 &   617 &      4.1E+05 \\
5 & 01:36:59.4 & +30:03:25.2 &     7.8 &       10.0 &       
        1.65 &       2.12 &      0.18 &     -375 &    -241 &  20.1 &   71.5 &      4.7E+04 \\       
        6 & 01:18:02.8 & +28:29:37.2 &    108.1 &       38.0 &     
        22.96 &       8.07 &      1.85 &     -370 &  -230 &     17.3 &   1.42E+04 &     9.3E+06 \\
       7 & 01:16:17.9 & +30:37:41.0 &    34.4 &       19.0 &     7.31 &       4.03 &      0.30 &     -383 &  -239 &       24.9 &   816 &      5.4E+05 
\enddata
\tablenotetext{a} {\scriptsize At a distance of 730 kpc.}
   \tablenotetext{b} {\scriptsize Located at the edge of the cube.}
\end{deluxetable}

\clearpage

\begin{figure}
\vspace{1in}
\includegraphics[angle=0, scale=0.75,viewport=8 80 670 720]{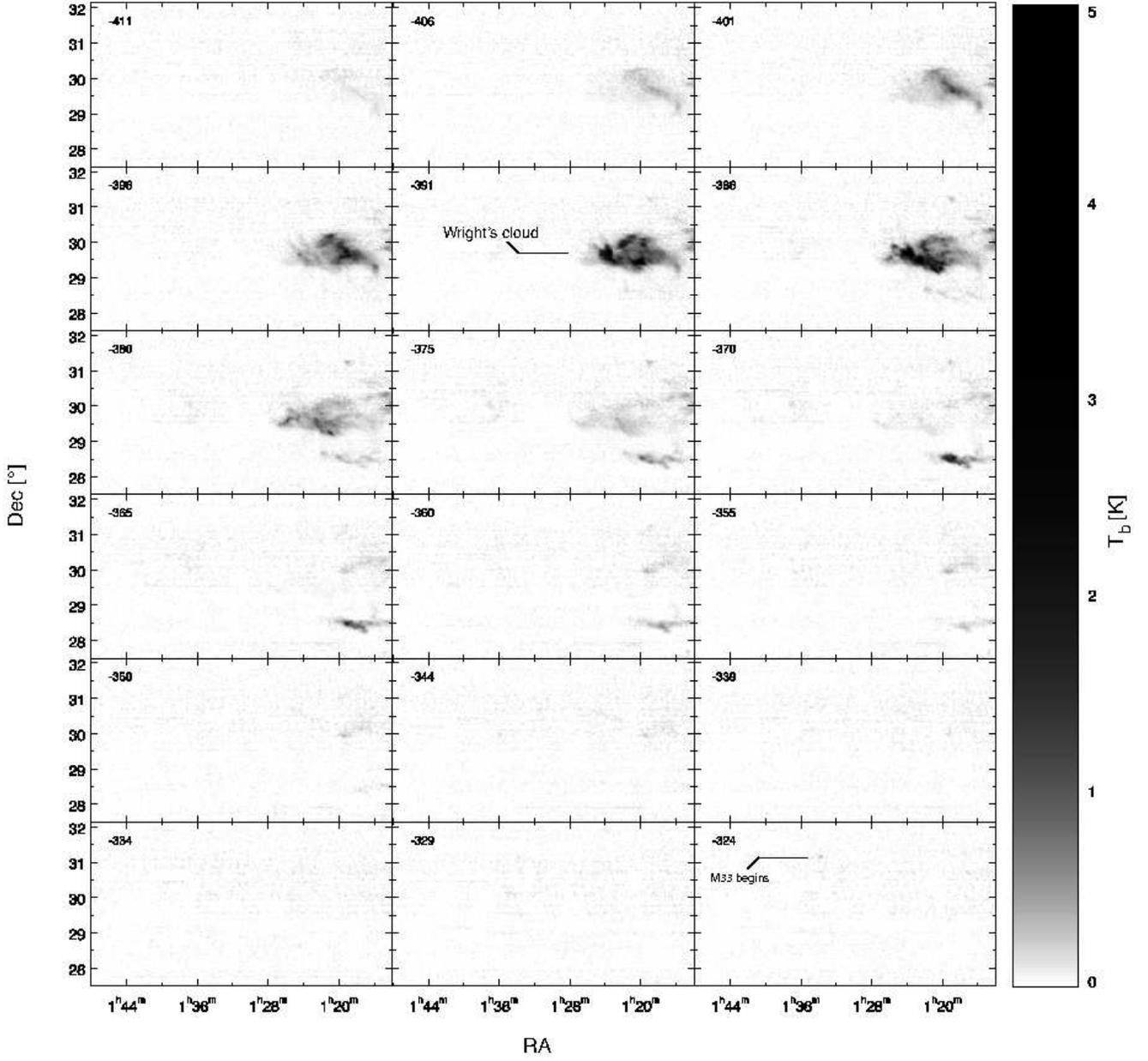}
\caption{Channel maps every 5 \kms\ from -411 to -46 \kms~showing the HI emission from Wright's Cloud, M33, and some Galactic emission.  The color bar at the right shows the intensity in Kelvin of
the greyscale and the velocity is labeled in the upper left.   Outside of
the channels that include Galactic emission, the 3$\sigma$ detection level is 0.1 K, or 10$^{18}$ \cm~per 5 \kms~channel.}  \label{hichans}
\end{figure}

\addtocounter{figure}{-1}
\begin{figure}
\vspace{1in}
\includegraphics[angle=0, scale=0.75,viewport=8 80 675 720]{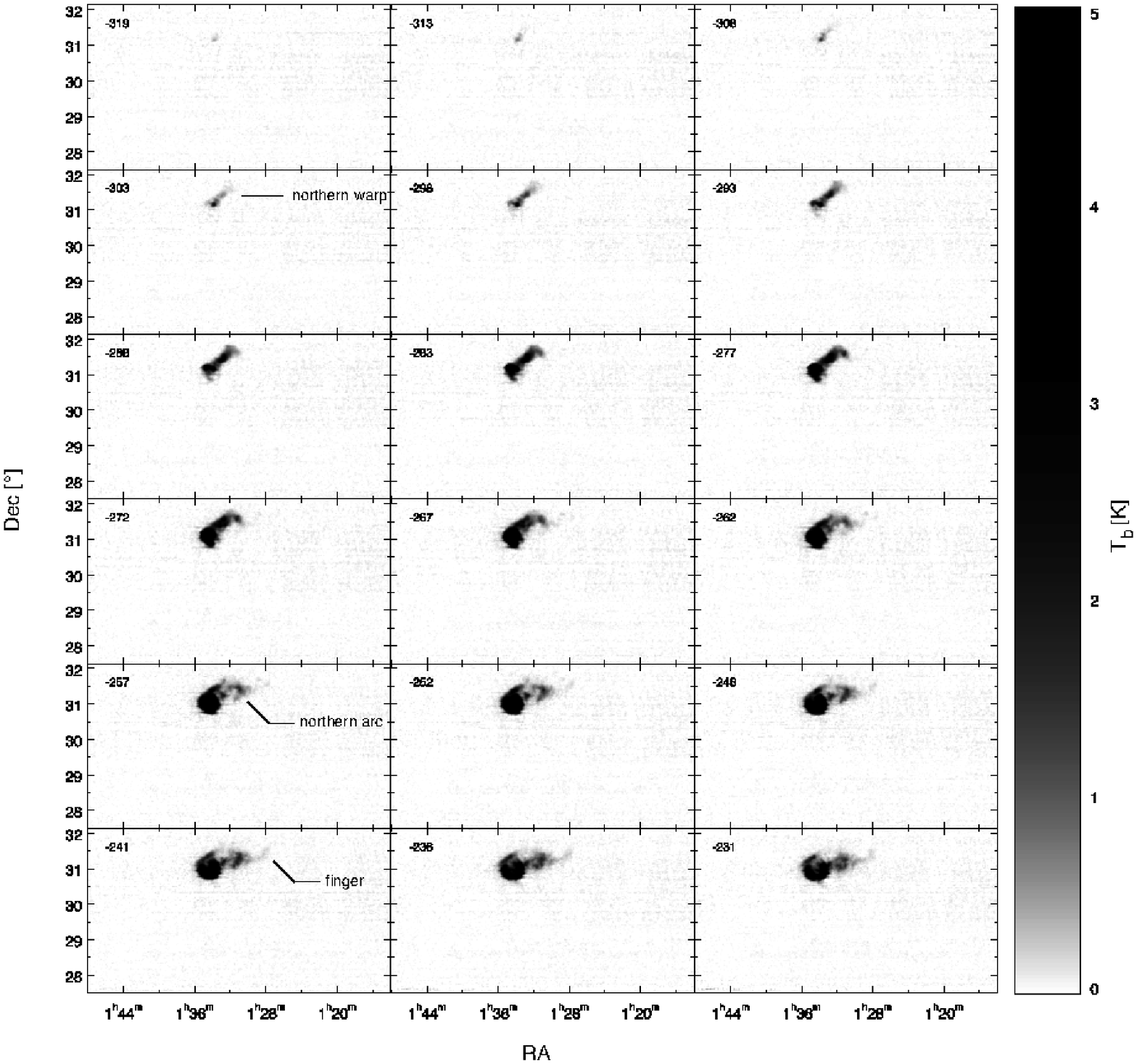}
\caption{(cont.)}
\end{figure}

\addtocounter{figure}{-1}
\begin{figure}
\vspace{1in}
\includegraphics[angle=0, scale=0.75,viewport=8 80 675 720]{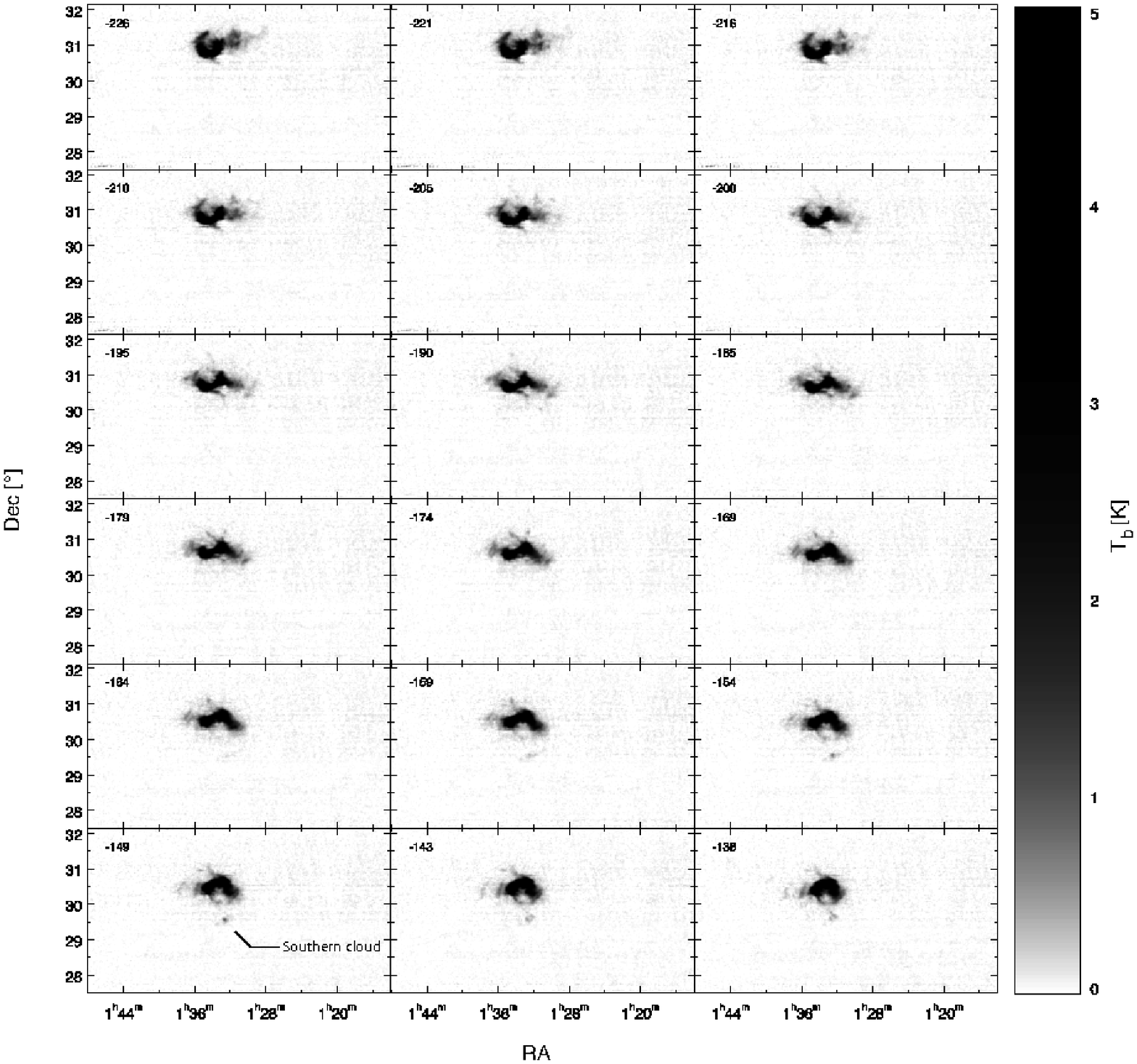}
\caption{(cont.)}
\end{figure}

\addtocounter{figure}{-1}
\begin{figure}
\vspace{1in}
\includegraphics[angle=0, scale=0.75,viewport=8 80 675 720]{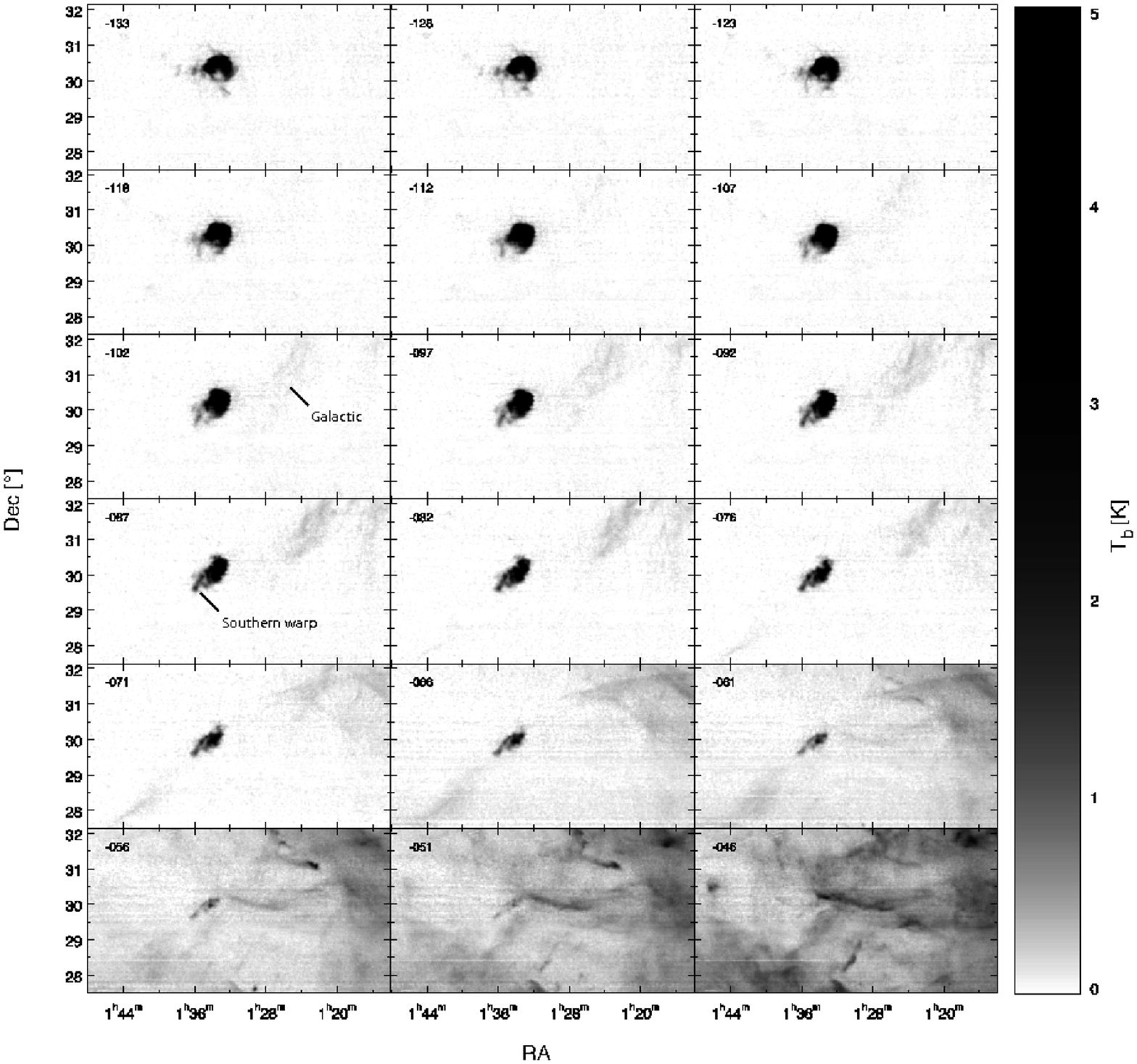}
\caption{(cont.)}
\end{figure}

\begin{figure}
\epsscale{0.95}
\plotone{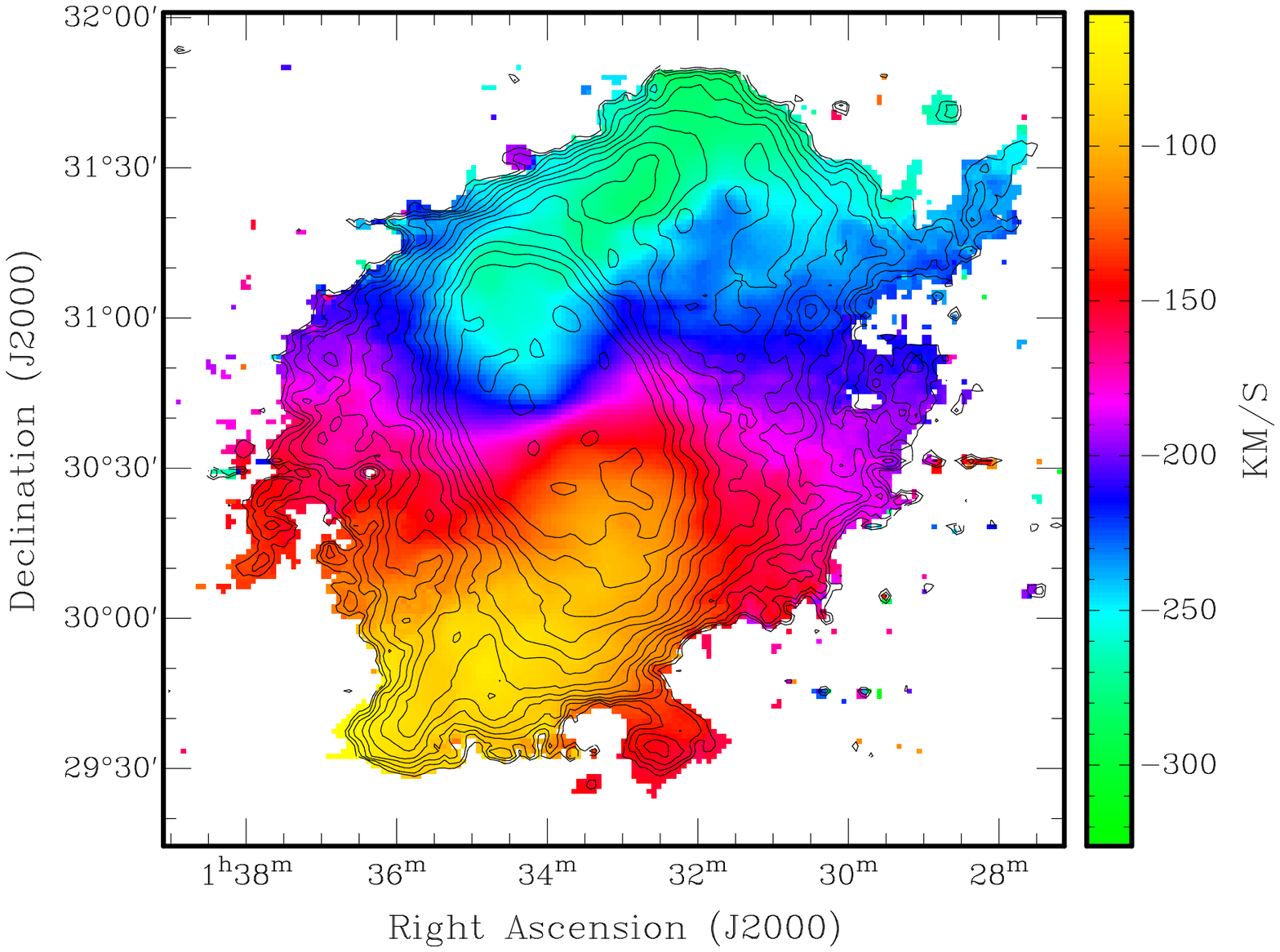}
\caption{The intensity weighted velocity (LSR) of the M33 gas with HI column density contours from Figure~\ref{hi} overlaid. } \label{hivel}
\end{figure}

\begin{figure}
\epsscale{0.95}
\plotone{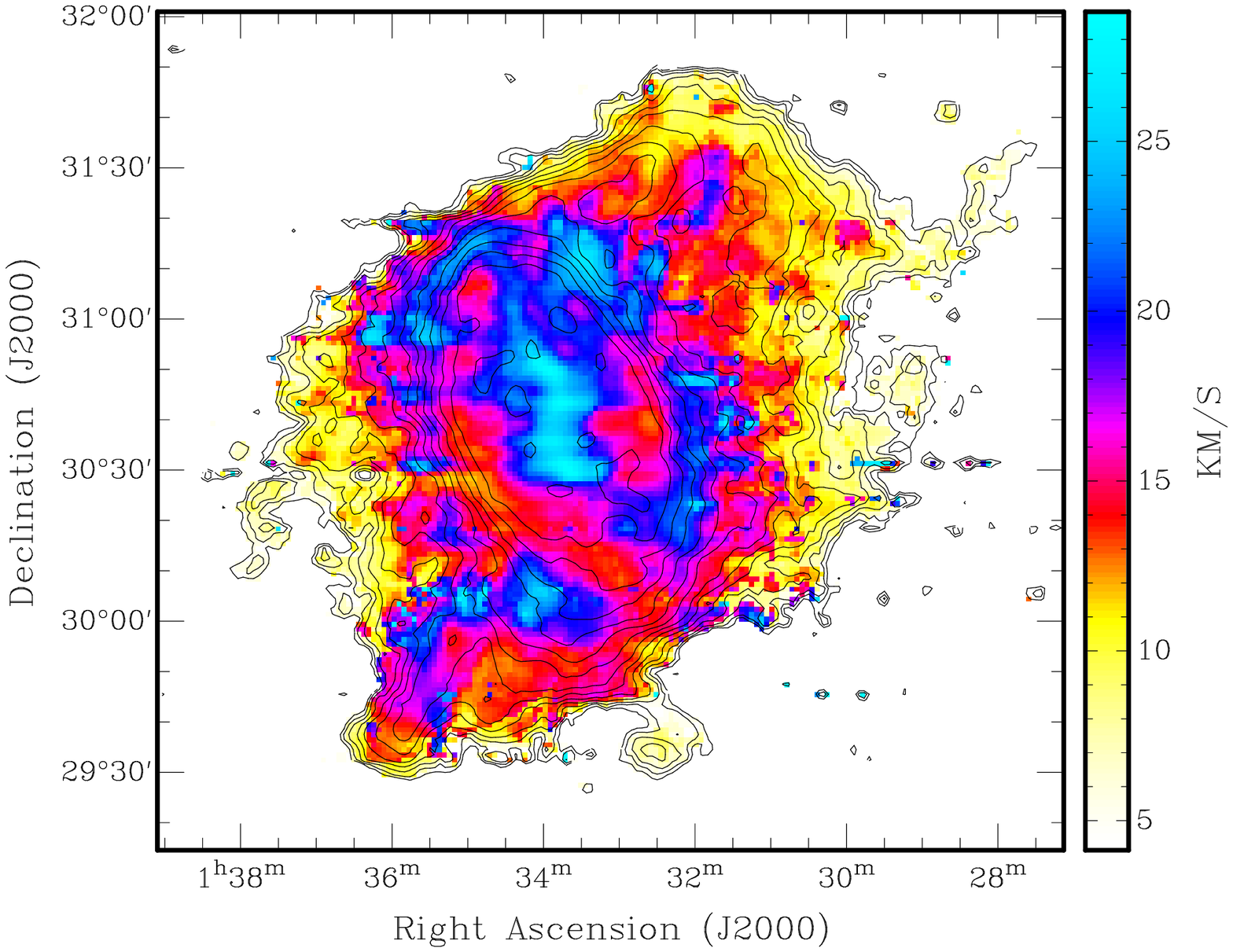}
\caption{The intensity weighted velocity dispersion map of M33 with HI column density contours from Figure~\ref{hi} overlaid.  Some of the scanning artifacts from the drift scans are more evident in this map.} \label{hiveldisp}
\end{figure}

\begin{figure}
\epsscale{0.9}
\plotone{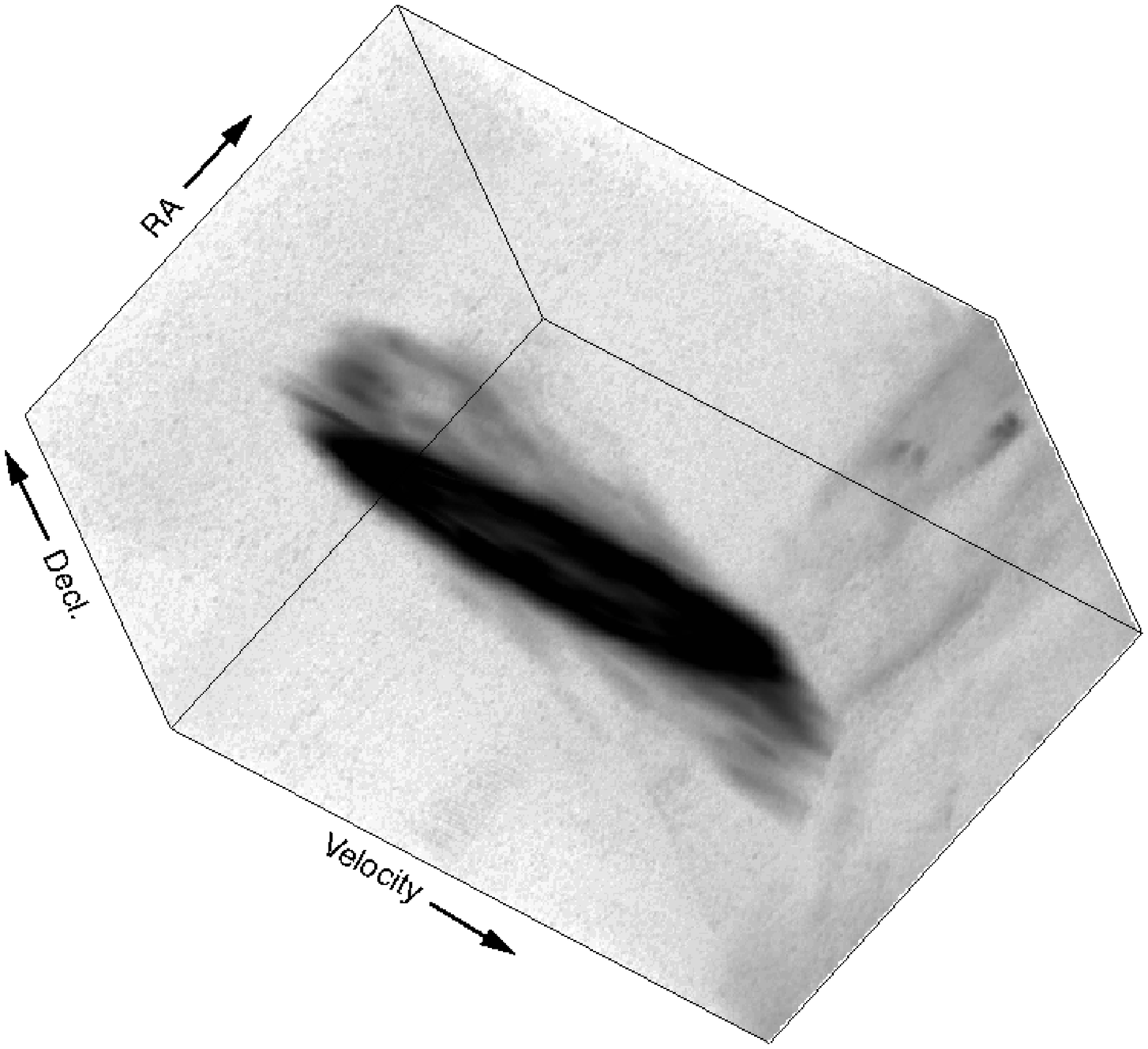}
\caption{Projection of the three-dimensional HI cube of M33 with the intensity contrast from 0 - 15 K on a log scale. Velocity extends along the bottom
axis and includes -360 to -52 \kms, therefore including some Galactic emission to the right.  } \label{xray}
\end{figure}

\begin{figure}
\vspace{0.2in}
\includegraphics[angle=0,scale=0.8]{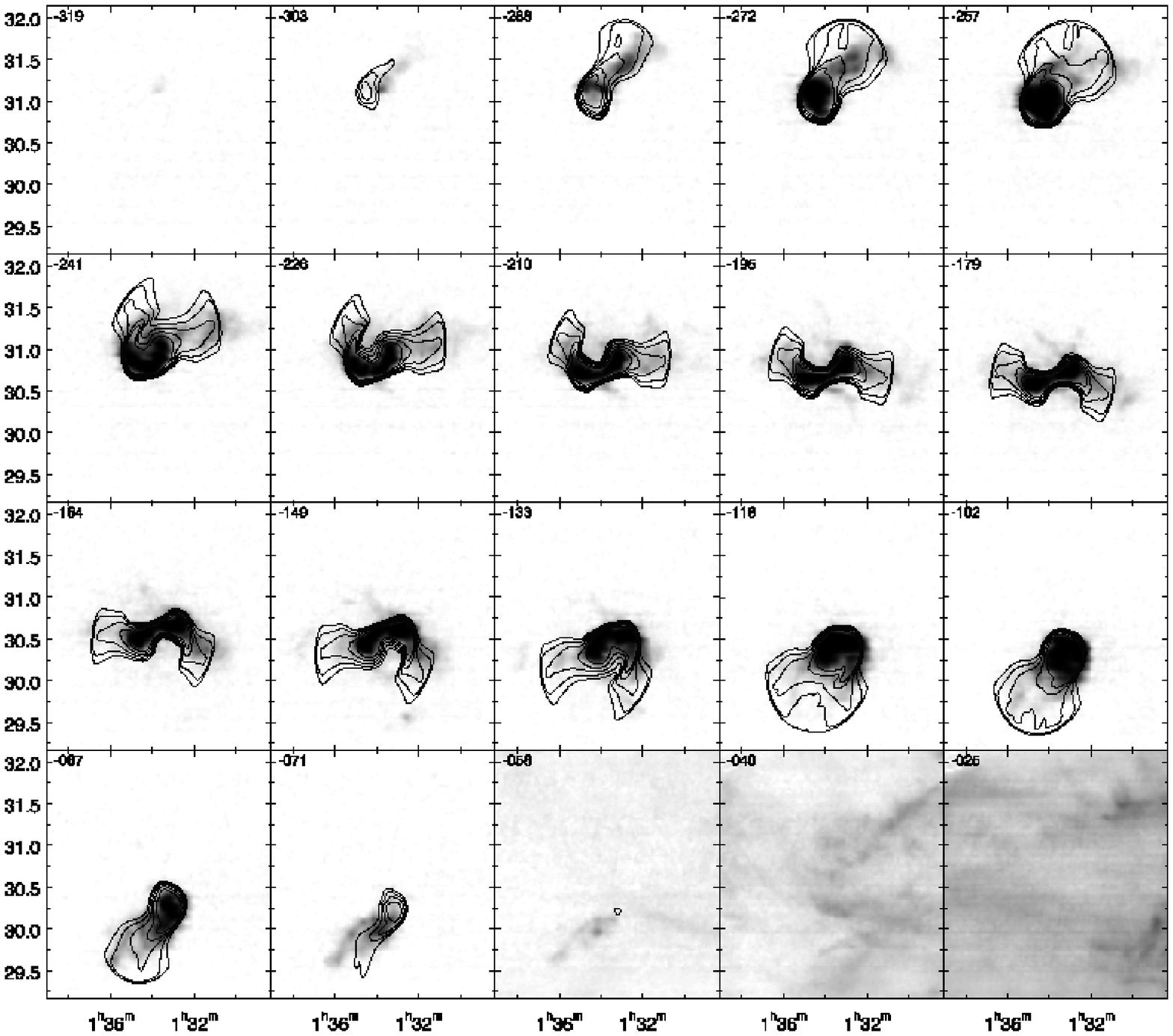}
\vspace{0.1in}
\caption{Channel maps showing the warped disk model in contours over the HI data incrementing every 15 \kms.  The grey scale for the HI data extends from 0 - 40 K on a log scale and the contours for the model are 0.2, 0.4, 0.8, 1.6, and 3.2 K.} \label{modchans}
\end{figure}

\begin{figure}
\epsscale{0.95}
\plotone{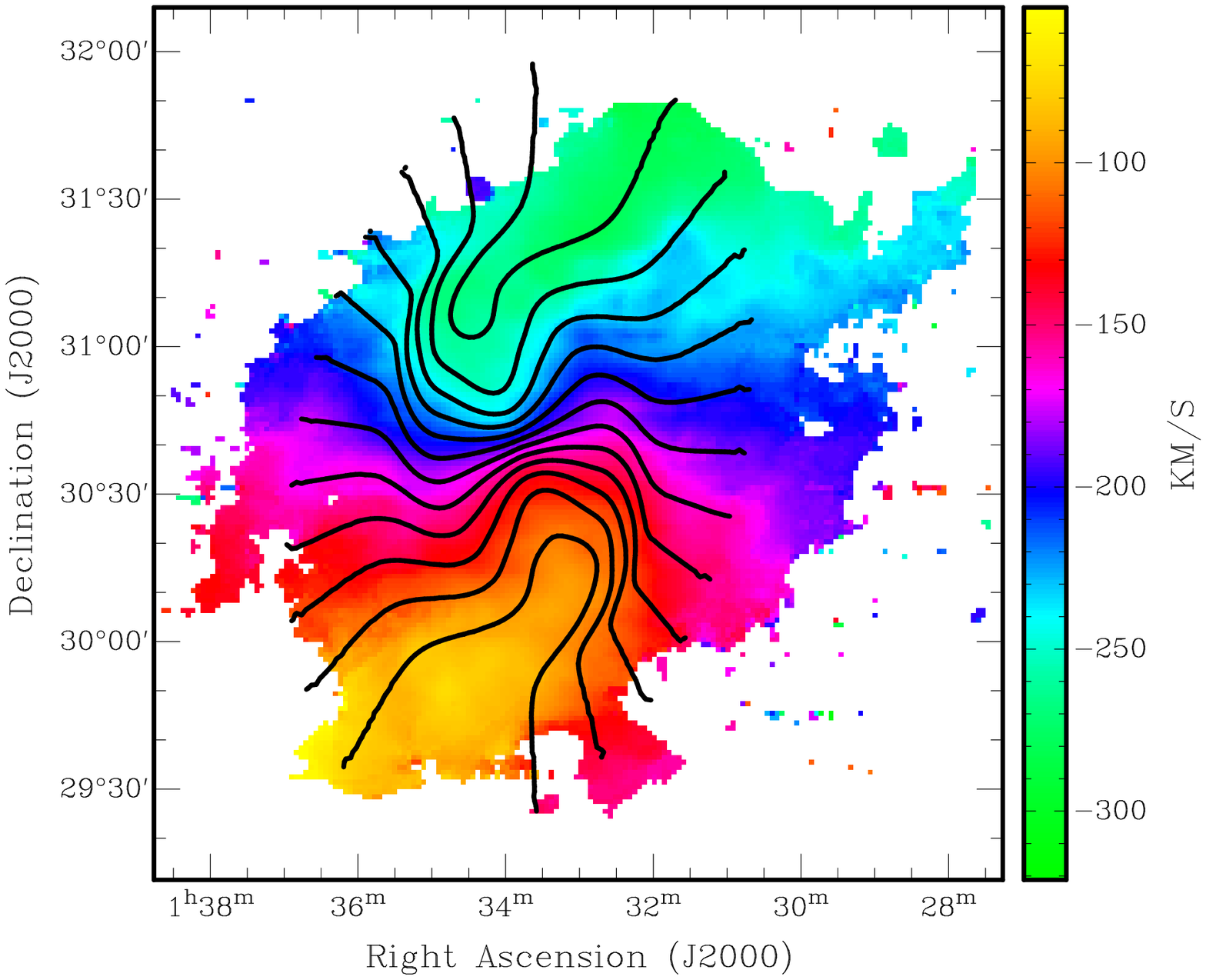}
\caption{The intensity weighted velocity of M33's HI (see Figure~\ref{hivel}) with the velocity contours of the warped disk model overlaid.  The contours start at -270 \kms at the top and extend to -75 \kms at the bottom incrementing in 15 \kms intervals.} \label{modvel}
\end{figure}

\begin{figure}
\vspace{0.3in}
\includegraphics[angle=0, viewport=0 0 515 250]{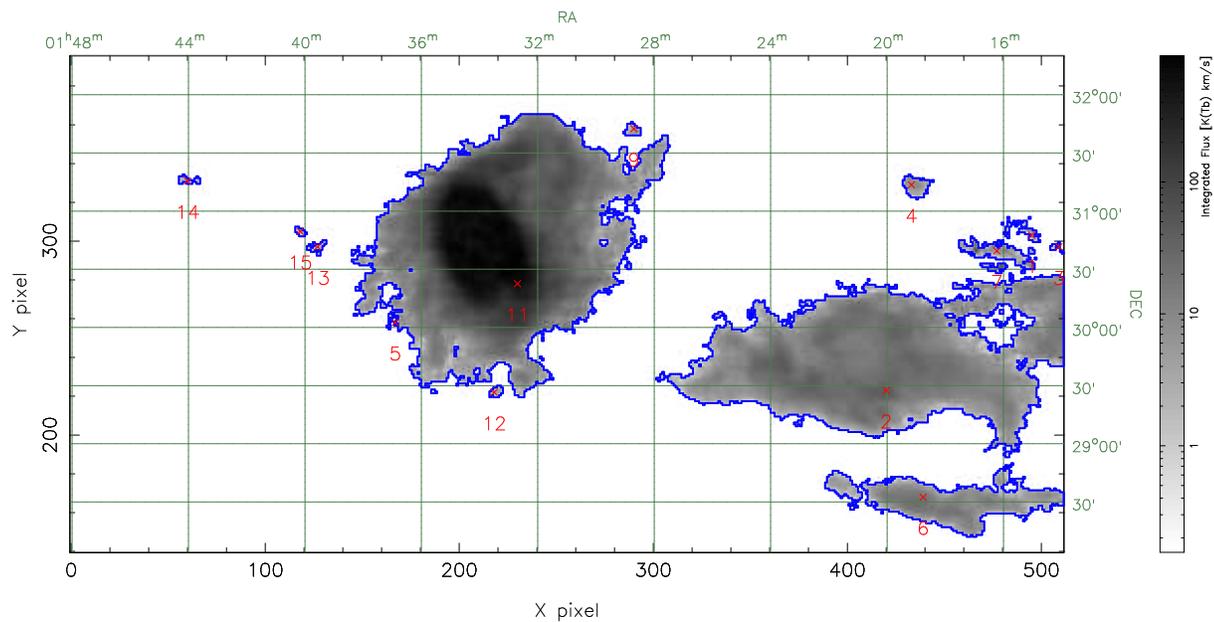}
\caption{The integrated intensity map with the objects found by Duchamp outlined and labeled (see Table 2).   The velocity range of the search was cut-off at -100 \kms, so M33 is not fully shown to the south.}\label{duchamp}
\end{figure}

\begin{figure}
\plotone{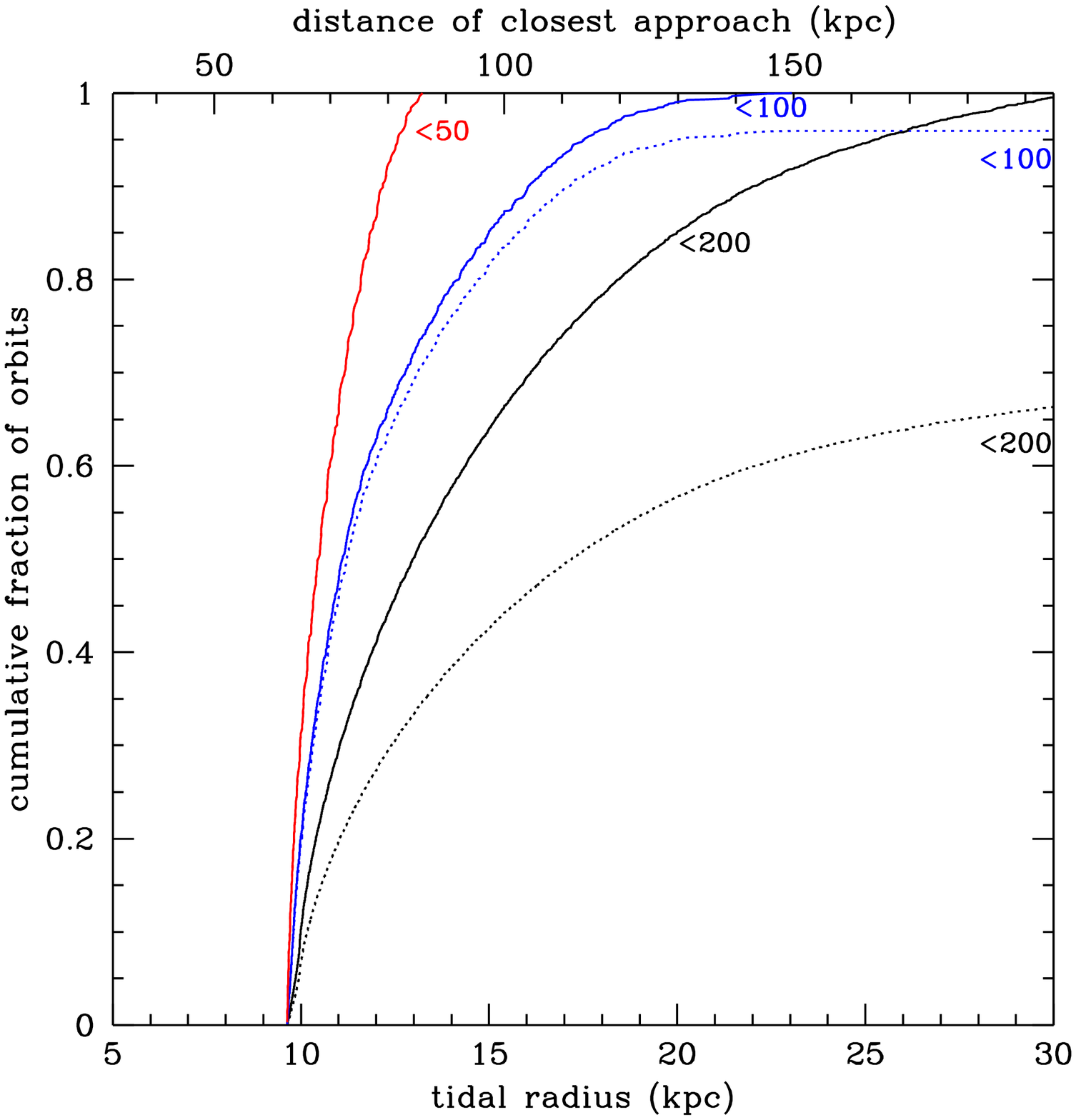}
\caption{Cumulative fraction of orbits that result in the tidal radius
for M33 noted on the bottom x-axis and and the distance of closest
approach between M31 and M33 (perigalacticon) noted on the top x-axis.
The three different pairs of lines show subsets of orbits with the maximum
tangential velocity of M31 restricted to be below 50, 100, and
200 \kms, respectively.  The solid lines are for orbits that had a
perigalacticon closer than M33's current separation ($\sim 200$ kpc),
and the dashed lines represent the total grid of orbits.  } \label{fig:sim1}
\end{figure}

\begin{figure}
\epsscale{1.1}
\plotone{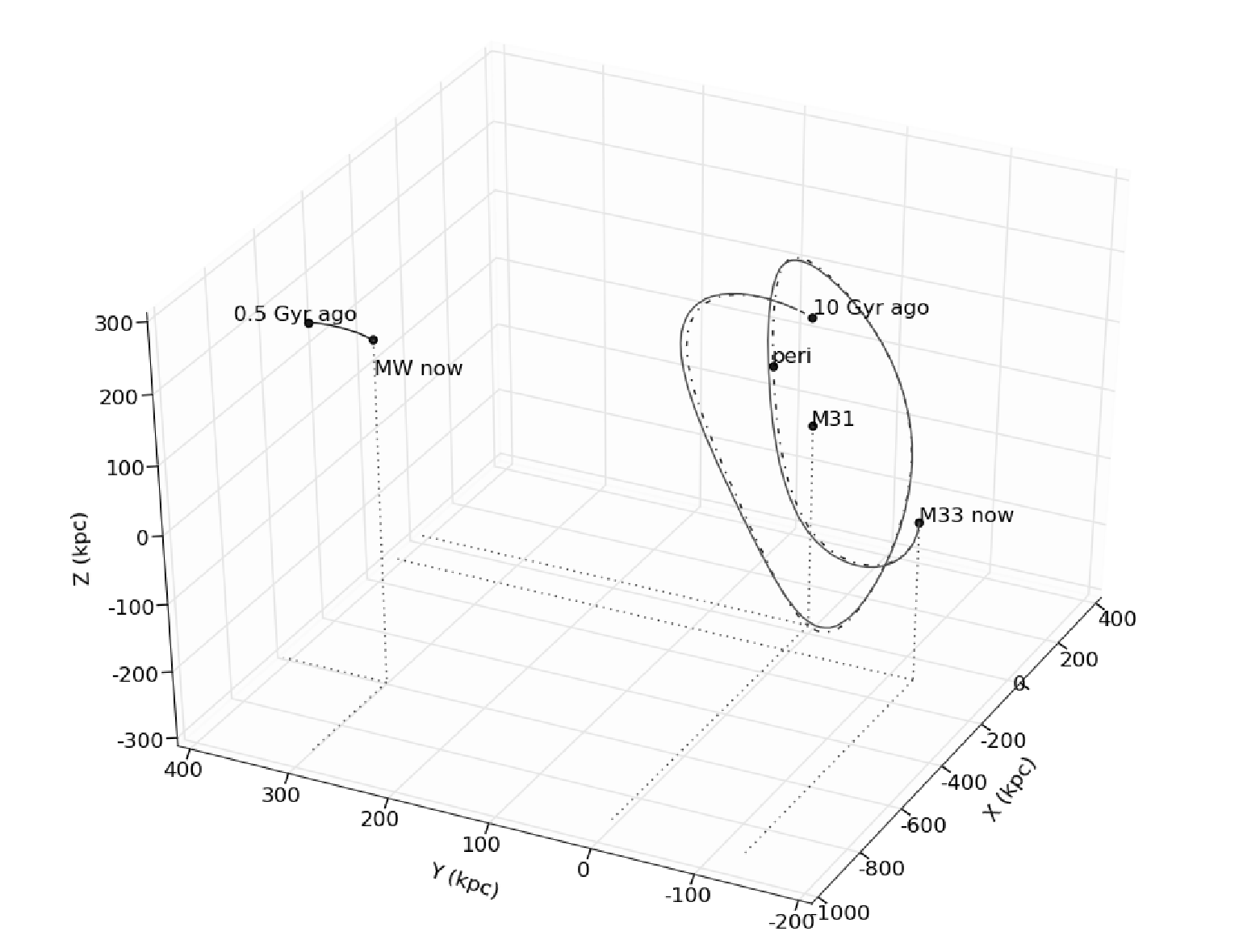}
\caption{A three-dimensional representation of a typical possible orbit of M33
shown in the M31 reference frame.  The solid line shows M33's orbit when
the Milky Way potential is included and the dashed line shows the orbit when
it is excluded.  'peri' marks the closest approach of M33 to M31 ($\sim10$ kpc) on
this orbit 1.6 Gyr ago.  The Milky Way moves outside the plotted box
after only $\approx 0.5$ Gyr in the past and therefore has little
influence on the orbit.\label{fig:m33orbit}}
\end{figure}

\begin{figure}
\includegraphics[angle=0,scale=0.9]{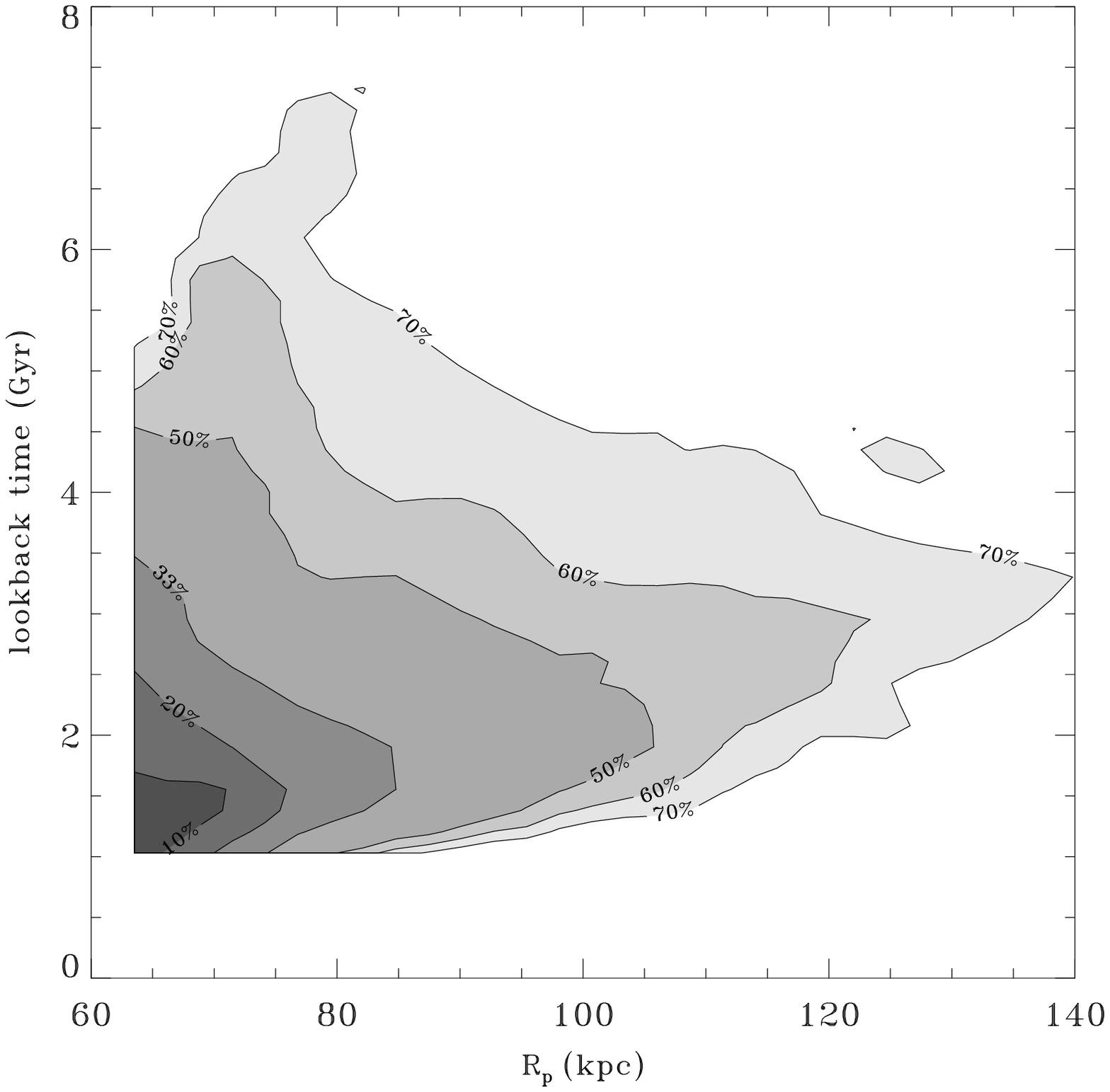}
\caption{The distribution of perigalactic distances and lookback times of the
closest approach between M33 and M31, for the full range of M31
tangential velocities up to 200 \kms.  The contours represent the
cumulative percentage of calculated orbits within the noted range
of perigalactica and lookback times.  The contours have abrupt left
and bottom sides because there are no orbits with $R_p < 63$ kpc and $t_p <
1$ Gyr.\label{fig:radtime}}

\end{figure}

\begin{figure}
\includegraphics[angle=0,scale=0.9]{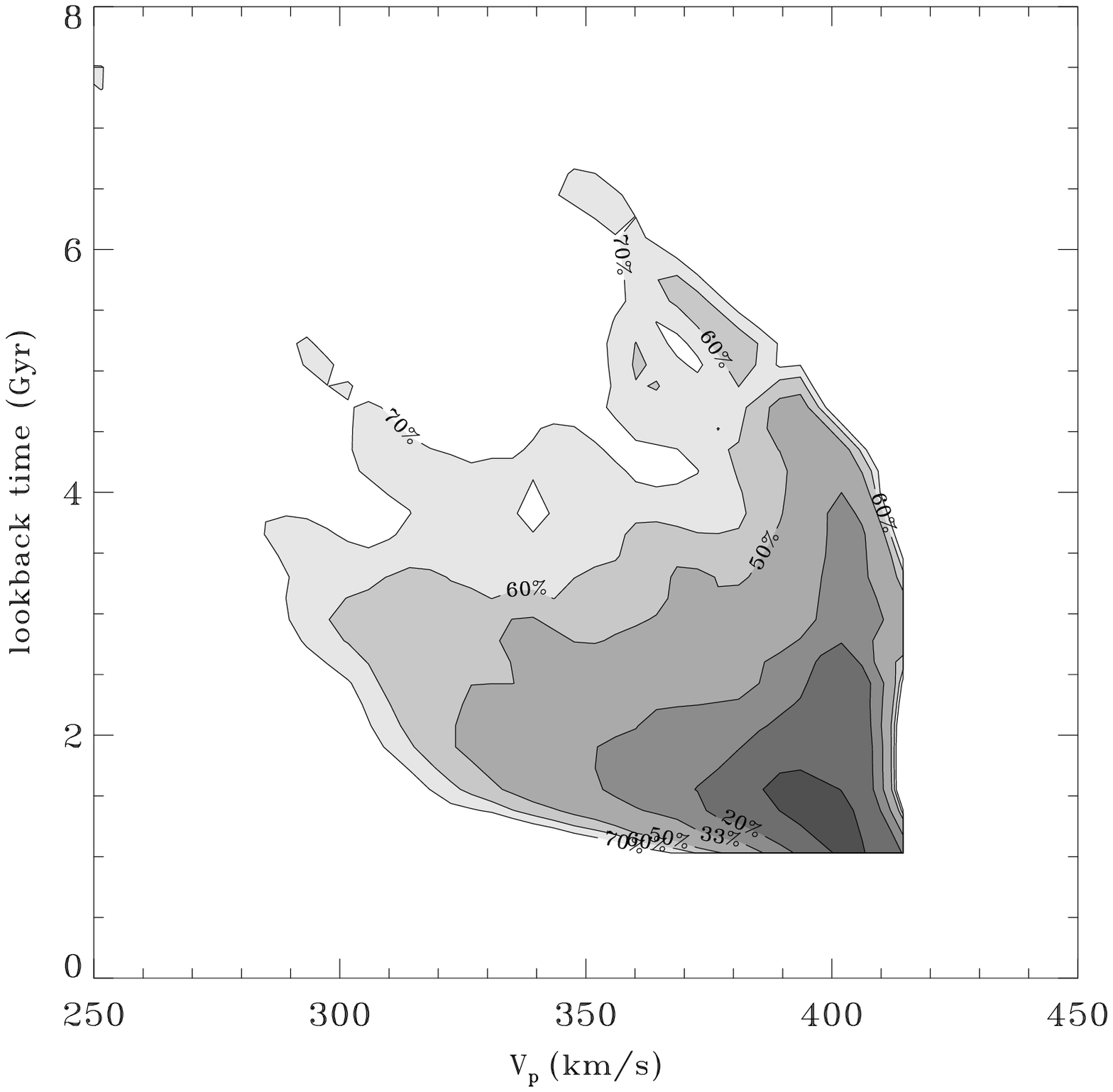}
\caption{The distribution of velocities and lookback times at the
closest approach between M33 and M31.  The contours represent the
cumulative percentage of calculated orbits within the noted range of
velocities and lookback times.\label{fig:veltime}}
\end{figure}

\end{document}